\documentclass[prd,aps,preprint,amsmath,amssymb,eshowkeys,nofootinbib]{revtex4-1}
\usepackage[mathscr]{euscript}
\usepackage{dcolumn}
\usepackage{multirow}
\usepackage{bm}
\usepackage{graphicx}
\usepackage{subfigure}
\usepackage{rotating}
\usepackage{amsmath,amssymb,amsthm}
\usepackage[colorlinks=true,linkcolor=red,urlcolor=green,citecolor=green]{hyperref}
\newcommand{\bea}{\begin{eqnarray}}
\newcommand{\eea}{\end{eqnarray}}
\newcommand{\beq}{\begin{equation}}
\newcommand{\eeq}{\end{equation}}

\def\/{\over}

\begin{document}

\title{Constant-roll inflation and primordial black holes within Barrow entropic framework}
\author{Qihong Huang$^{1}$\footnote{Corresponding author: huangqihongzynu@163.com}, Li-Yang Chen$^{2}$, He Huang$^{3}$, Bing Xu$^{4}$ and Kaituo Zhang$^{5}$}
\affiliation{
$^1$ School of Physics and Electronic Science, Zunyi Normal University, Zunyi, Guizhou 563006, China\\
$^2$ College of Physics and Electronic Engineering, Chengdu Normal University, Chengdu, Sichuan 611130, China\\
$^3$ Institute of Applied Mechanics, Zhejiang University, Hangzhou, Zhejiang 310058, China\\
$^4$ School of Electrical and Electronic Engineering, Anhui Science and Technology University, Bengbu, Anhui 233030, China\\
$^5$ Department of Physics, Anhui Normal University, Wuhu, Anhui 241000, China
}

\begin{abstract}
In this paper, starting from the modified Einstein field equations, we derive the modified scalar spectral index $n_{s}$ and the modified tensor-to-scalar ratio $r$ in Barrow entropy model, calculate their values for the power-law, periodic, and hilltop potential models, constrain the model parameter $\delta$ and the potential parameter using Planck 2018 data, and find that increasing $\delta$ causes a significant decrease in $r$. Then, we calculate the primordial curvature perturbation power spectra, primordial black hole (PBH) abundance, and scalar induced gravitational waves (SIGWs) for these models, finding PBH mass of approximately $10^{-12} M_{\odot}$, PBH abundance nearly $0.98$, and the peak frequencies of SIGWs on the order $10^{-3} \mathrm{Hz}$, indicating that these models not only generate sufficient PBHs which can contribute one-third of the dark matter content but could also be detected by next-generation missions such as LISA, Taiji, and TianQin. Subsequently, we analyze the evolution of PBHs and find that when the effective equation of state parameter evolves from $1/3$ to $-1/3$, the accretion mass increases to approximately $10^{2}M_{i}$, while the temperature of the PBHs decreases from $10^{4}K$ to $10^{2}K$, suggesting that PBHs exist and are detectable today.
\end{abstract}

\maketitle

\section{Introduction}

Inflation, a short period of an exponential accelerating expansion before the radiation dominated era, is the currently widely accepted paradigm of modern cosmology~\cite{Guth1981, Linde1982}. Inflation not only addresses the puzzles of the standard hot Big Bang cosmology but also provides an explanation for the quantum origin of the Cosmic Microwave Background (CMB) temperature anisotropies and the Large-Scale Structure~\cite{Mukhanov1981, Lewis2000, Bernardeau2002}. In general, a scalar field named as the inflaton field drives the exponential accelerating expansion of the universe during the inflationary epoch. The mechanism of inflation is based on the generation of small quantum fluctuations in the inflaton field. The small quantum fluctuations are amplified in physical scale during the inflationary epoch and lead to a Gaussian, scale-invariant and adiabatic primordial density perturbations~\cite{Weinberg2008}. This information is encoded into the primordial scalar power spectrum described by the scalar spectral index $n_{s}$, which is constrained by the Planck 2018 results~\cite{Planck2020}. In addition, inflation also predicts the generation of tensor perturbations as primordial gravitational waves described by the tensor-to-scalar ratio $r$~\cite{Maggiore2018}. Usually, the dynamics of inflation is based on the slow-roll approximation, in which the scalar potential is chosen to be nearly flat so that the scalar field can slowly roll down this potential. Once the minimum value of this potential is reached, inflation ends. This scenario is called slow-roll inflation~\cite{Linde1982, Noh2001, Weinberg2008}, which can be measured by the parameters $\epsilon_{1}=-\dot{H}/H^{2}$ and $\epsilon_{2}=\ddot{\phi}/H \dot{\phi}$. For the slow-roll inflation, the small values of the parameters $\epsilon_{1} \ll 1$ and $\epsilon_{2} \ll 1$ are required. If the scalar potential is assumed to be extremely flat, the second condition becomes $\epsilon_{2}=-3$ which corresponds to the ultra-slow-roll condition~\cite{Martin2013, Dimopoulos2017, Pattison2018}. In the ultra-slow-roll inflation, the curvature perturbations are not frozen at the super Hubble scales thus leading to non-Gaussianities. Furthermore, when the second condition $\epsilon_{2}$ is generalized to be constant, a more generalized scenario named as constant-roll inflation was proposed~\cite{Motohashi2015, Gao2017, Gao2018, Yi2018, Anguelova2018, Cicciarella2018, Guerrero2020}, in which the rate of acceleration and velocity of the inflaton field are constants. The constant-roll inflation exhibits distinct dynamical features that enrich inflationary physics by extending beyond the slow-roll paradigm. It delivers exact analytical solutions inaccessible under slow-roll approximations, exhibits superhorizon evolution of curvature perturbations in specific parameter regimes, and demonstrates parameter-dependent attractor stability~\cite{Motohashi2015}. These features broaden the scope of testable inflation scenarios, particularly through viable models such as natural or hilltop inflation, enabling new probes of early-universe dynamics. When the constant-roll inflation was proposed, it drawn widespread attention and was widely studied in lots of theories, such as $f(R)$ gravity~\cite{Nojiri2017a, Motohashi2017}, $f(R,\phi)$ gravity~\cite{Panda2023}, $f(Q,T)$ gravity~\cite{Bourakadi2023}, scalar-tensor gravity~\cite{Motohashi2019}, brane gravity~\cite{Mohammadi2020}, Dirac-Born-Infeld theory~\cite{Lahiri2022}, Galilean model~\cite{Herrera2023}, non-minimally coupled model~\cite{Karam2018, Shokri2021}, non-minimally derivative coupling~\cite{Liu2024}, quintessential model~\cite{Shokri2022}, holographic dark energy~\cite{Mohammadi2022, Nojiri2023}, Tsallis holographic dark energy~\cite{Keskin2023}, and so on.

After inflation ended, the primordial inhomogeneities in the primordial power spectrum on small scales re-entered the Hubble horizon during the radiation dominated era, and a significant amount of primordial black holes (PBHs) can be formed as a result of the gravitational collapse~\cite{Hawking1971, Carr1974} if the amplitude of the primordial power spectrum produced during inflation is sufficiently large. After PBHs were proposed, they were found that they could be a candidate of dark matter~\cite{Chapline1975} and were reconsidered~\cite{Bird2016, Sasaki2016} after black holes mergers were detected by LIGO~\cite{Abbott2016}. PBHs can offer us an opportunity to explore the physics of the early universe and may play some important roles in cosmology~\cite{Bagui2025}. Thus, PBHs were widely discussed in some modified gravity theories, such as scalar-tensor gravity~\cite{Barrow1996}, $f(T)$ gravity~\cite{Bhadra2013}, Brans-Dicke gravity~\cite{Aliferis2021}, $f(Q)$ gravity~\cite{Chanda2022}, teleparallel gravity~\cite{Bourakadi2022}, $f(Q,T)$ gravity~\cite{Bourakadi2023}, and studied in some inflationary models including slow-roll inflation~\cite{Carr1993, Ivanov1994, Yokoyama1998, Bellido2017, Pi2018, Biagetti2018, Fu2020, Davies2022, Lin2020, Yi2021, Gao2021, Gao2021a, Yi2021a, Wu2021, Wang2024, Wang2024a, Chen2024, Solbi2024, Frolovsky2025, Afzal2025, Choudhury2025, Cook2023}, ultra-slow-roll inflation~\cite{Di2018, Fu2019, Ballesteros2020, Liu2021, Figueroa2022, Zhai2022, Chen2022, Zhai2023, Choudhury2024, Su2025} and constant-roll inflation~\cite{Motohashi2020, Tomberg2023, Bourakadi2023}. PBHs also have attracted lots of attention since they may constitute part or all of dark matter~\cite{Clesse2015, Kawasaki2016, Carr2016, Inomata2017, Inomata2018, Carr2018, Germani2019, Kusenko2020, Pacheco2020, Calabrese2022, Pacheco2023, Flores2023, Dike2023}, they may play a role in the synthesis of heavy elements~\cite{Fuller2017, Keith2020, Calza2025} and could be responsible for some astrophysical phenomena, such as, seeding supermassive black holes~\cite{Kawasaki2012, Clesse2015, Carr2018, Kusenko2020}, seeding galaxies~\cite{Clesse2015, Carr2018}, explaining the gravitational wave signals observed by the LIGO detectors~\cite{Inomata2017a, Kusenko2020, Wang2023}. The existence of PBHs formed in the early universe remains an open question. PBHs with masses smaller than $5 \times 10^{14}g$ would have evaporated by Hawking radiation by now~\cite{Page1976}. When PBHs are considered as dark matter, studies have suggested various possible mass ranges~\cite{Inomata2017, Kuhnel2017, Carr2017, Dike2023, DAgostino2023}, though these proposals differ significantly in scale. Consequently, the allowed mass window for PBHs appears broad, but a more precise range requires further investigation. However, recent analyses have converged on the "asteroid mass" range of $10^{17}-10^{23}g$, within which PBHs could potentially explain the entire dark matter~\cite{Carr2016, Green2024}.

The holographic dark energy is one competitive candidate for dark energy~\cite{Hsu2004, Horvat2004, Li2004}, which is used to explain the late time acceleration of the universe, and is proposed based on the holographic principle stating that the entropy of a system is scaled on its surface area~\cite{Witten1998, Bousso2002}. The cornerstone of holographic dark energy is the horizon entropy, and different horizon entropy will result in different holographic dark energy models. Combining the holographic principle with Barrow entropy~\cite{Barrow2020a}, Barrow holographic dark energy with different IR cutoffs have been proposed~\cite{Saridakis2020, Anagnostopoulos2020a, Srivastava2021a, Sheykhi2021, Oliveros2022} and subsequently widely studied in theories~\cite{Huang2021, Adhikary2021, Mamon2021, Rani2021, Luciano2022, Nojiri2022, Paul2022, Luciano2023, Boulkaboul2023, Pankaj2023, Ghaffari2023} and observations~\cite{Luciano2023, Asghari2022, Jusufi2022, Barrow2021, Anagnostopoulos2020}. Based on the Barrow entropy and Clausius relation, the modified gravitational field equations in Barrow entropy model were proposed~\cite{Asghari2022}. These extensive studies have precisely constrained the key parameter of the Barrow entropy model $\delta$, which has been determined to have a small value consistent with observational results~\cite{Asghari2022, Jusufi2022, Barrow2021, Anagnostopoulos2020, Luciano2023}. To preserve the Big Bang Nucleosynthesis(BBN) epoch, $\delta$ must be within the bound $\delta \leq 1.4 \times 10^{-4}$~\cite{Barrow2021}, which provides the tightest constraint for $\delta$. However, while $\delta$ is tightly constrained by BBN and other late-time observations, its role during the inflationary epoch characterized by extreme energy scales and quantum gravitational effects remain largely unexplored within Barrow cosmology. Recently, the constant-roll inflation was studied in holographic dark energy model~\cite{Mohammadi2022} and Tsallis holographic dark energy model~\cite{Keskin2023}, the results show that the constant-roll inflation can be realized in these models under some conditions. So, whether the scalar spectral index $n_s$ and the tensor-to-scalar ratio $r$ depend on $\delta$ in the Barrow entropy model during inflation; how the modified model deviates from standard cosmology when $\delta$ assumes a not negligible values; whether observationally viable constant-roll inflation can be realized in this model; and if achievable, whether it can produce PBH abundances compatible with current constraints.

This paper has two main objectives: to investigate constant-roll inflation and to explore the formation and evolution of PBHs within the Barrow entropic framework. The paper is organized as follows: In Section II, we briefly review the modified Friedmann equation within Barrow entropy model. In Section III, we investigate constant-roll inflation within Barrow entropy model. In Section IV, we explore the formation and evolution of PBHs within Barrow entropy model. Finally, our main conclusions are shown in Section V.

\section{Modified Friedmann Equations}

In a homogeneous and isotropic Friedmann-Robertson-Walker universe described by the Friedmann-Lema\^{\i}tre-Robertson-Walker metric
\beq
ds^{2}=-dt^{2}+ a(t)^2 \gamma_{ij} dx^{i} dx^{j},
\eeq
where $a(t)$ denotes the scale factor, and $\gamma_{ij}$ represents the metric on the three-sphere
\beq
\gamma_{ij} dx^{i} dx^{j}=\frac{dr^{2}}{1-K r^{2}}+r^{2}(d\theta^2+\sin^{2}\theta d\phi^2).
\eeq
Here, $K$ is the spatial curvature. 

Combining the Barrow entropy of the black hole, the first law of thermodynamics, and the apparent horizon as the IR cutoff, the Friedmann equation within the Barrow entropy model is modified as~\cite{Saridakis2020a, Barrow2021, Sheykhi2021, Leon2021, Sheykhi2022, Sheykhi2023, Luciano2023, Sheykhi2023a, Salehi2023, Motaghi2024, Sheykhi2025, Keskin2025}
\beq\label{H1}
\Big(H^{2}+\frac{K}{a^{2}}\Big)^{1-\frac{\delta}{2}}=\frac{8\pi G_{eff}}{3}\rho_{\phi},
\eeq
where $\delta$ satisfies the relation $0 \leq \delta \leq 1$ and stands for the amount of the quantum-gravitational deformation effects~\cite{Barrow2020}, and $G_{eff}$ represents the effective Newtonian gravitational constant as
\beq
G_{eff}=\frac{A_{0}}{4}\Big(\frac{2-\delta}{2+\delta}\Big)\Big(\frac{A_{0}}{4\pi}\Big)^{\frac{\delta}{2}}.
\eeq
In the case $\delta \rightarrow 0$, the area law of entropy is recovered, and we have $A_{0} \rightarrow 4G$. As a result, $G_{eff}\rightarrow G$ and the standard Friedmann equation is obtained. In addition, Eq.~(\ref{H1}) can also be obtained from the modified gravitational field equations within the Barrow entropy model~\cite{Asghari2022}
\beq\label{MGFE}
R_{\mu\nu} A^{\frac{\delta}{2}} - \nabla_{\mu}\nabla_{\nu} A^{\frac{\delta}{2}} - \frac{1}{2} R A^{\frac{\delta}{2}} g_{\mu\nu} + \Box A^{\frac{\delta}{2}} g_{\mu\nu} = \frac{4\pi}{2+\delta} A^{1+\frac{\delta}{2}}_{0} T_{\mu\nu},
\eeq
in which for a flat spacetime, one have
\beq
A^{\frac{\delta}{2}} = (4\pi)^{\frac{\delta}{2}} H^{-\delta}.
\eeq

We consider the matter contents of the early universe as a scalar field, and the energy density and the pressure take the form
\beq
\rho_{\phi}=\frac{1}{2}\dot{\phi}^{2}+V(\phi), \qquad p_{\phi}=\frac{1}{2}\dot{\phi}^{2}-V(\phi),
\eeq
which satisfies the continuity equation
\beq\label{s}
\dot{\rho}_{\phi}+3H(\rho_{\phi}+p_{\phi})=0.
\eeq
It can be written as the Klein-Gordon equation
\beq\label{s2}
\ddot{\phi}+3H\dot{\phi}+V_{\phi}=0,
\eeq
where $V_{\phi} = \frac{\partial V}{\partial \phi}$. Combing Eqs.~(\ref{H1}) and ~(\ref{s}), the second Friedmann equation can be derived as~\cite{Sheykhi2021}
\beq\label{H2}
(2-\delta)\Big(1+\frac{\dot{H}}{H^{2}}\Big)\Big(H^{2}+\frac{K}{a^{2}}\Big)^{-\frac{\delta}{2}}H^{2}+(1+\delta)\Big(H^{2}+\frac{K}{a^{2}}\Big)^{1-\frac{\delta}{2}}=-8\pi G_{eff} p_{\phi}.
\eeq

\section{Constant-roll Inflation}

In this section, we will discuss the constant-roll inflation for the Friedmann equations given by Eqs.~(\ref{H1}) and ~(\ref{H2}) in a flat universe. During inflation, the universe is characterized by the scalar spectral index $n_{s}$ and the tensor-to-scalar ratio $r$. To obtain the expressions for $n_{s}$ and $r$ within the Barrow entropy model, we derive the Mukhanov-Sasaki equation using the method in Ref.~\cite{Garriga1999, Mukhanov1992}. This method is often applied to derive the curved Mukhanov-Sasaki equation~\cite{Handley2019, Thavanesan2021, Shumaylov2022, HuangQ2022, HuangQ2023, HuangQ2023a}. By considering the perturbed line element in linear perturbation theory under the Newtonian gauge, the perturbed metric can be expressed as
\beq\label{dsp}
ds^{2}=-(1+2\Psi)dt^{2}+ a(t)^2 [(1+2\Phi)\gamma_{ij}+2h_{ij}] dx^{i} dx^{j}.
\eeq
where $\Psi$ and $\Phi$ are scalar perturbations, and $h_{ij}$ is a transverse-traceless tensor perturbation. Then, the perturbed (00) and (0i) components of the modified gravitational field equations ~(\ref{MGFE}) can be written as~\cite{Asghari2022}
\bea
&& H^{-\delta}\Big[ -\frac{1}{a^{2}}\nabla^{2}\Phi+3H(H \Phi+\dot{\Phi})-\frac{3}{2}\delta\Big(2\dot{H}\Phi+\frac{\dot{H}}{H}\dot{\Phi}\Big) \Big]=-4\pi G \delta T^{0}_{0}, \\
&& H^{-\delta}\Big[ \nabla_{i}(-H \Phi-\dot{\Phi})+\frac{1}{2}\delta \Big(\frac{\dot{H}}{H}\Big)\nabla_{i}\Phi \Big]=-4\pi G \delta T^{0}_{0}.
\eea
After simplifying the above equations, we obtain
\bea
&& \Big(\frac{\delta\phi}{\dot{\phi}}\Big)^{\cdot}=\Phi+\Big[\frac{1}{a^{2}}\nabla^{2}\Phi+\frac{3}{2}\delta\Big(\dot{H}\Phi+\frac{\dot{H}}{H}\dot{\Phi}\Big)\Big]\frac{H^{-\delta}}{4\pi G \dot{\phi}^{2}},\label{pH1}\\
&& (a \Phi)^{\cdot}=4\pi G a\Big(\dot{\phi}^{2}\frac{\delta\phi}{\dot{\phi}}\Big)H^{\delta}+\frac{1}{2}\delta\Big(\frac{\dot{H}}{H}\Big)a\Phi.\label{pH2}
\eea
Introducing the new variables $\xi$ and $\zeta$ defined as~\cite{Garriga1999}
\bea
&& \Phi a=4\pi G H \xi,\\
&& \frac{\delta\phi}{\dot{\phi}}=\frac{\zeta}{H}-\frac{4\pi G}{a}\xi,
\eea
Eqs.~(\ref{pH1}) and ~(\ref{pH2}) can be written as
\bea
&& \dot{\xi}=\frac{a \dot{\phi}^{2}}{H^{2-\delta}}\zeta,\label{pm1}\\
&& \dot{\zeta} \simeq \frac{H^{2-\delta}}{a \dot{\phi}^{2}}\frac{\Delta \xi}{a^{2}}.\label{pm2}
\eea 
Here, $\Delta=\nabla^{2}$ and the slow-roll condition $\frac{\dot{\phi}^{2}}{2}\ll1$ is used. Following Ref.~\cite{Garriga1999}, we obtain the first order action which reproduces the equations of motion ~(\ref{pm1}) and ~(\ref{pm2}) is
\beq
S =\int\Big( \xi \dot{\zeta}-\frac{1}{2}\frac{H^{2-\delta}}{a^{3}\dot{\phi}^{2}}\xi \Delta \xi+\frac{1}{2}\frac{a \dot{\phi}^{2}}{H^{2-\delta}}\zeta^{2} \Big)dt d^{3}x.
\eeq
Introducing $v=z \zeta$, the above equation becomes
\beq
S =\frac{1}{2}\int z^{2}(\zeta'^{2}+\zeta \Delta \zeta)d\eta d^{3}x=\frac{1}{2}\int \Big(v'^{2}+v \Delta v+\frac{z''}{z}v^{2}\Big)d\eta d^{3}x \label{pv}
\eeq
with
\beq
z^{2}=\frac{a^{2-\delta} \phi'^{2}}{\mathcal{H}^{2-\delta}},\label{z2}
\eeq
where a prime $'$ denotes the derivative with respect to the conformal time $\eta$, and $\mathcal{H}=\frac{a'(\eta)}{a(\eta)}$. Then, the equation for the variable $v$ immediately follows from Eq.~(\ref{pv}) and takes the form
\beq
v''-\Delta v-\frac{z''}{z}v=0,
\eeq
which gives the Mukhanov-Sasaki equation
\beq
v_{k}''+\Big(k^{2}-\frac{z''}{z}\Big)v_{k}=0,\label{MSE}
\eeq
with
\beq
\frac{z''}{z}=\frac{\nu^{2}-\frac{1}{4}}{\eta^{2}}, \quad \nu^{2} \simeq \frac{9}{4} + \Big( 6-\frac{3}{2}\delta \Big)\epsilon_{1} + 3\epsilon_{2}.
\eeq
Here, $\epsilon_{1}$ and $\epsilon_{2}$ are the first and second slow-roll parameters, respectively, defined as
\beq
\epsilon_{1}=-\frac{\dot{H}}{H^{2}}, \qquad \epsilon_{2}=\frac{\ddot{\phi}}{H \dot{\phi}}.
\eeq
Considering the Bunch–Davies vacuum as the initial condition, the solution to the Mukhanov-Sasaki equation~(\ref{MSE}) becomes
\beq
v_{k} = \frac{1}{\sqrt{2k}} e^{i(\nu-\frac{1}{2})\frac{\pi}{2}} 2^{(\nu-\frac{3}{2})} \frac{\Gamma(\nu)}{\Gamma(\frac{3}{2})} (-k \eta)^{(\frac{1}{2}-\nu)},
\eeq
which has the same form as that in standard slow-roll inflation. Then, the power spectrum of the primordial curvature perturbations $\mathcal{P}_{\mathcal{R}}(k)$ is given as
\beq\label{PRS}
\mathcal{P}_{\mathcal{R}}(k)=\frac{k^{3}}{2\pi^{2}} \Big| \frac{v_{k}}{z} \Big|^{2} = 2^{2\nu-3} \Big( \frac{\Gamma(\nu)}{\Gamma(\frac{3}{2})} \Big)^{2} \Big( \frac{H}{\dot{\phi}} \Big)^{2} \Big( \frac{H}{2\pi} \Big)^{2} \Big( \frac{k}{aH} \Big)^{3-2\nu} H^{-\delta} \Big |_{k=k_{*}}.
\eeq
Here, $k_{*}=0.002\mathrm{Mpc}^{-1}$ corresponds to the CMB scale. Thus, the scalar spectral index $n_{s}$ can be calculated as
\beq\label{nsm}
n_{s} = 1 + \frac{d \ln \mathcal{P}_{\mathcal{R}}}{d \ln k} \Big |_{k=k_{*}} = 4 - 2\nu \simeq 1-(4-\delta)\epsilon_{1}-2\epsilon_{2}.
\eeq

Following ~\cite{Dodelson2021}, combining the modified gravitational field equations ~(\ref{MGFE}) with the perturbed metric ~(\ref{dsp}), we can obtain the tensor perturbation equation
\beq\label{tpe1}
h''_{ij} + k^{2} h_{ij} + \Big[ 2\mathcal{H} + \delta\Big( \mathcal{H}-\frac{\mathcal{H}'}{\mathcal{H}} \Big) \Big] h'_{ij}= 0.
\eeq
Performing a Fourier transform on the tensor perturbation $h_{ij}$, we obtain~\cite{Bagui2025}
\beq
h_{ij}(\eta,\vec{x})=\int \frac{d^{3}k}{(2\pi)^{3}} \sum\limits_{T=+,\times} h^{T}_{k}(\eta) e^{T}_{ij}(\hat{k}) e^{i \vec{k} \cdot \vec{x}},
\eeq
where $h^{T}_{k}$ are the two helicity modes, and $e^{T}_{ij}$ is the transverse-traceless polarization tensor. After introducing the variables $v^{T}_{k}$ and $z_{T}$
\beq
v^{T}_{k} = z_{T} h^{T}_{k}, \quad z_{T} = \frac{a}{\sqrt{16\pi G_{eff}}}\Big( \frac{a}{\mathcal{H}} \Big)^{\frac{\delta}{2}},
\eeq
the tensor perturbation equation ~(\ref{tpe1}) can be written as
\beq
v^{T\prime\prime}_{k} + \Big( k^{2} - \frac{z''_{T}}{z_{T}} \Big) v^{T}_{k}=0,
\eeq
with
\beq
\frac{z''_{T}}{z_{T}} = \frac{\nu^{2}_{T}-\frac{1}{4}}{\eta^{2}}, \qquad \nu^{2}_{T} \simeq \frac{9}{4}+3\Big(1+\frac{1}{2}\delta\Big)\epsilon_{1}.
\eeq
Then, the power spectrum of the tensor perturbations $\mathcal{P}_{\mathcal{T}}(k)$ is given by
\beq\label{PRT}
\mathcal{P}_{\mathcal{T}}(k) = 2\frac{k^{3}}{\pi^{2}} \Big| \frac{v^{T}_{k}}{z_{T}} \Big|^{2} = 64 \pi G_{eff} \times 2^{2\nu_{s}-3} \Big( \frac{\Gamma(\nu_{T})}{\Gamma(\frac{3}{2})} \Big)^{2} \Big( \frac{H}{2\pi} \Big)^{2} \Big( \frac{k}{aH} \Big)^{3-2\nu_{T}} H^{-\delta} \Big |_{k=k_{*}},
\eeq
and the tensor spectral index $n_{T}$ can be calculated as
\beq
n_{T} = \frac{d \ln \mathcal{P}_{\mathcal{T}}}{d \ln k} \Big |_{k=k_{*}} = 3-2\nu_{T} \simeq -(2+\delta)\epsilon_{1}.
\eeq
Combining Eqs.~(\ref{PRS}) and ~(\ref{PRT}), the tensor-to-scalar ratio $r$ is obtained
\beq\label{rm}
r=\frac{\mathcal{P}_{\mathcal{T}}}{\mathcal{P}_{\mathcal{R}}} \simeq 8(2-\delta)\epsilon_{1} H^{\delta}.
\eeq

In order not to spoil the BBN epoch, the Barrow exponent $\delta$ requires satisfying $\delta \leq 1.4 \times 10^{-4}$~\cite{Barrow2021}. Thus, under this constraint, the scalar spectral index ~(\ref{nsm}) and the tensor-to-scalar ratio ~(\ref{rm}) reduce to the one in general relativity~\cite{Nojiri2017, Hwang2005}
\beq
n_{s} \simeq 1-4\epsilon_{1}-2\epsilon_{2}, \qquad r=16\epsilon_{1},
\eeq
which has been used to analyze inflation in Tsallis holographic dark energy~\cite{Keskin2023} and Barrow holographic dark energy~\cite{Keskin2025}. 

For the slow-roll inflation, $\epsilon_{1}\ll 1$ and $\epsilon_{2}\ll 1$ are required. While in the constant-roll inflation, imposing $\frac{\dot{\phi}^{2}}{2}\ll1$, only $\epsilon_{1}\ll 1$ is required to occur in the inflation, and the other constant-roll condition is given by
\beq\label{csc}
\ddot{\phi}=\gamma H \dot{\phi}.
\eeq
Here, $\gamma$ is a dimensionless real parameter. For $\gamma=-3$, the ultra-slow-roll condition, which has a flat potential $V_{\phi}=0$, is recovered. And the slow-roll condition is obtained for $\gamma=0$. In the flat universe, considering $\frac{\dot{\phi}^{2}}{2}\ll1$, Eqs.~(\ref{H1}), ~(\ref{H2}) and ~(\ref{s2}) can be written as
\bea
&& H^{2}\simeq \Big( \frac{8\pi G_{eff}}{3}\Big)^{\frac{2}{2-\delta}} V^{\frac{2}{2-\delta}},\label{wr1}\\
&& \dot{H}\simeq -\frac{3\dot{\phi}^{2}}{2(2-\delta)} \Big( \frac{8\pi G_{eff}}{3}\Big)^{\frac{2}{2-\delta}} V^{\frac{\delta}{2-\delta}},\label{wr2}\\
&& \dot{\phi}=-\frac{V_{\phi}}{(\gamma+3)H}.\label{wr3}
\eea
Then, the constant-roll parameters become
\beq\label{crp}
\epsilon_{1}=\frac{3\dot{\phi}^{2}}{2(2-\delta)V}, \qquad \epsilon_{2}=\gamma.
\eeq

Similar to the slow-roll inflation occuring at the horizon crossing point, we also consider inflation begins at the horizon crossing point. Thus, the e-folds number $N$, which determines the amount of inflation, is given as
\beq
N=\int^{t_{end}}_{t_{*}} H dt,
\eeq
where $t_{*}$ and $t_{end}$ represent the horizon crossing time and the end of inflation time, respectively. Rewriting the e-folds number $N$ according to the scalar field, we obtain
\beq
N=\int^{\phi_{end}}_{\phi_{*}} \frac{H}{\dot{\phi}} d\phi.
\eeq
To analyze the relations among the e-folds number $N$, the scalar spectral index $n_s$, and the tensor-to-scalar ratio $r$, we focus on the three inflationary potentials inherent in constant-roll inflation: power-law potentials, periodic potentials, and hilltop potentials~\cite{Motohashi2015}.  

\subsection{Power-law potential}

Since power-law inflation with $V \sim e^{\phi}$ is excluded due to the tensor-to-scalar ratio $ r \approx 0.28 $~\cite{Motohashi2015}, we begin with the simplest power-law inflationary potential, which takes the form
\beq\label{Vp}
V=V_{0}\phi^{n},
\eeq
where $V_{0}$ and $n$ are positive parameters. This class of potentials includes the simplest chaotic inflationary models~\cite{Linde1983} where inflation begins at large field values, with applications spanning slow-roll inflation~\cite{Bassett2006, Planck2020, Tahmasebzadeh2016}, reheating~\cite{Goswami2018} and PBHs~\cite{Gao2021a}. At the end of inflation, the first constant parameter $\epsilon_{1}$ satisfies $\epsilon_{1}(\phi_{end})\simeq 1$ which gives
\beq
\phi_{end}=\Bigg[ \Bigg( \frac{8\pi G_{eff}}{3} \Bigg)^{\frac{2}{2-\delta}} \frac{2(2-\delta)(\gamma+3)^{2}}{3 n^{2}} V^{-\frac{\delta}{\delta-2}}_{0} \Bigg]^{\frac{\delta-2}{4-2\delta+n \delta}}.
\eeq
Here, Eqs. ~(\ref{wr1}), ~(\ref{wr3}) and ~(\ref{crp}) are taken into consideration. Then, the e-folds number can be written according $\phi_{end}$ and $\phi_{*}$
\beq
N=\Big( \frac{8\pi G_{eff}}{3} \Big)^{\frac{2}{2-\delta}} \frac{\gamma+3}{n} V^{\frac{\delta}{2-\delta}}_{0} \frac{\delta-2}{4-2\delta+n \delta} \Big( \phi_{end}^{-\frac{4-2\delta+n \delta}{\delta-2}} - \phi_{*}^{-\frac{4-2\delta+n \delta}{\delta-2}} \Big).
\eeq
Solving the above equation, we obtain the value of the scalar field at the horizon crossing $\phi_{*}$
\beq
\phi_{*}=\Bigg[ \Big( \frac{8\pi G_{eff}}{3} \Big)^{-\frac{2}{2-\delta}} V^{-\frac{\delta}{2-\delta}}_{0} \Big( \frac{3n^{2}}{2(2-\delta)(\gamma+3)^{2}}-\frac{N n}{\gamma+3} \frac{4-2\delta+n \delta}{\delta-2} \Big) \Bigg]^{-\frac{\delta-2}{4-2\delta+n \delta}}.
\eeq
Using this result, the first constant-roll parameter $\epsilon_{1}$ can be written as
\beq
\epsilon_{1}=\frac{3n}{3n+2(\gamma+3)[4+(n-2)\delta]N}.
\eeq

Then, the scalar spectral index $n_s$ and the tensor-to-scalar ratio $r$ become
\bea
&& n_{s}=1-\frac{3(4-\delta)n}{3n+2(\gamma+3)[4+(n-2)\delta]N}-2\gamma,\label{ns}\\
&& r=\frac{24(2-\delta)n}{3n+2(\gamma+3)[4+(n-2)\delta]N} H^{\delta},\label{r}
\eea
which show that the parameters $n_s$ and $r$ depend on the power-term $n$ of the power-law potential, the constant-roll parameter $\gamma$, the parameter of the Barrow entropy model $\delta$ and the e-folds number $N$.

\begin{figure*}[htp]
\begin{center}
\includegraphics[width=0.45\textwidth]{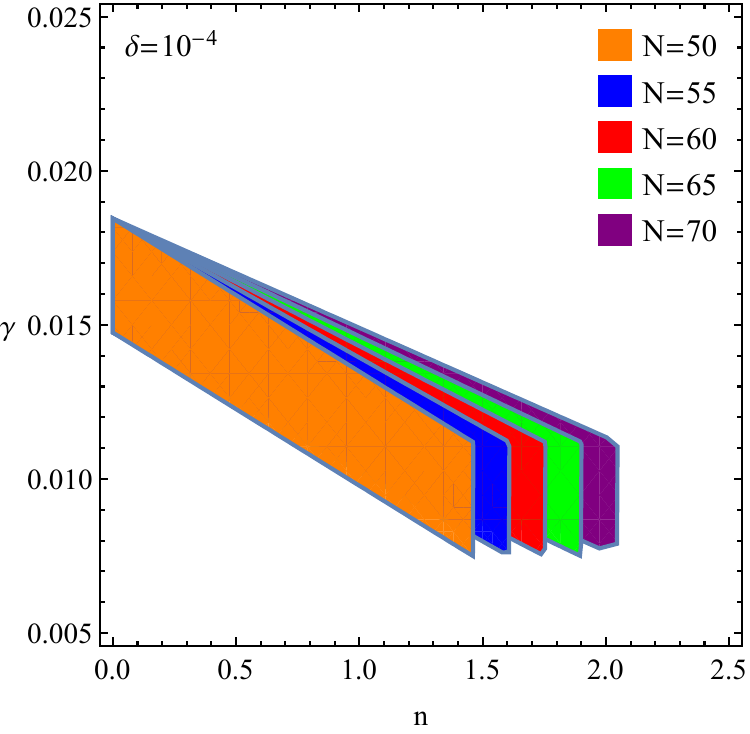}
\includegraphics[width=0.45\textwidth]{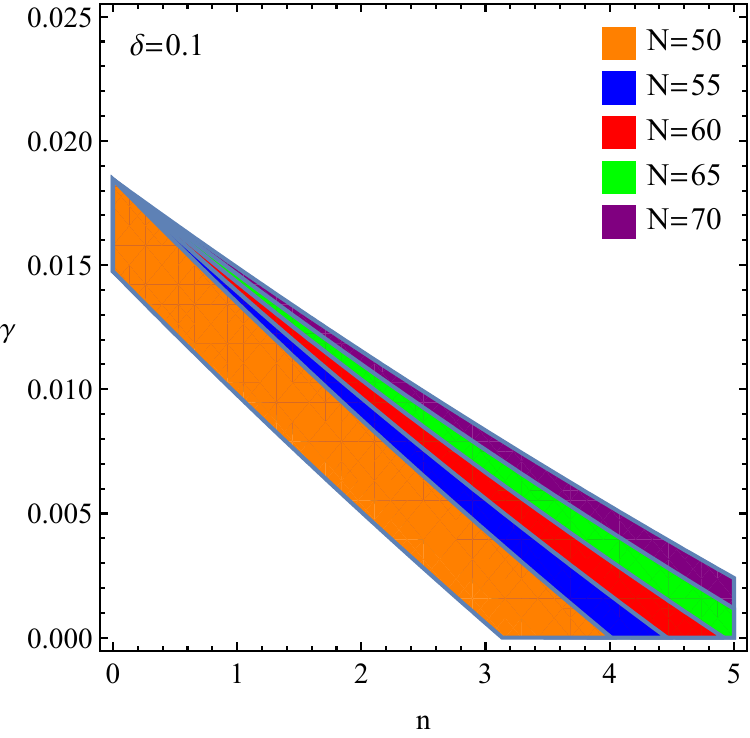}
\caption{\label{Fig1} Prediction regions of the power-law potential ~(\ref{Vp}) in $\gamma-n$ plane.}
\end{center}
\end{figure*}

To determine the value of the power-term $n$ in the power-law potential $V=V_{0}\phi^{n}$ and match our results with the observations, we adopt the constraints $r_{0.002}<0.058$ and $n_{s}=0.9668 \pm 0.0037$ from Planck TT,TE,EE + lowE + lensing + BK15 + BAO~\cite{Planck2020a}. For model parameters, we consider $\gamma$ in the range $0<\gamma<0.025$, while $\delta$ is assigned a small value consistent with observational results~\cite{Asghari2022, Jusufi2022, Barrow2021, Anagnostopoulos2020, Luciano2023}. To preserve the Big Bang Nucleosynthesis epoch, $\delta$ must be within the bound $\delta \leq 1.4 \times 10^{-4}$~\cite{Barrow2021}, which provides the tightest constraint for $\delta$. To analyze deviations of Barrow entropy model from the standard cosmology, we also explore non-negligible values of $\delta$ beyond observational bounds. Then, using Eqs. ~(\ref{ns}) and ~(\ref{r}), we have plotted the prediction regions in $\gamma-n$ plane with $\delta=10^{-4}$ and $\delta=0.1$, and with the number of e-folds $N$ ranging from $50$ to $70$ in Fig.~(\ref{Fig1}). These figures show that the value of $\gamma$ is very small, with $n$ determined by both $\gamma$ and the e-folds number $N$. Additionally, a larger value of $\delta$ expands the range of the $\gamma-n$ plane. For $\delta = 10^{-4}$, $n<1.46$ is required, while $n<2.61$ is required for $\delta = 0.1$.

\begin{figure*}[htp]
\begin{center}
\includegraphics[width=0.45\textwidth]{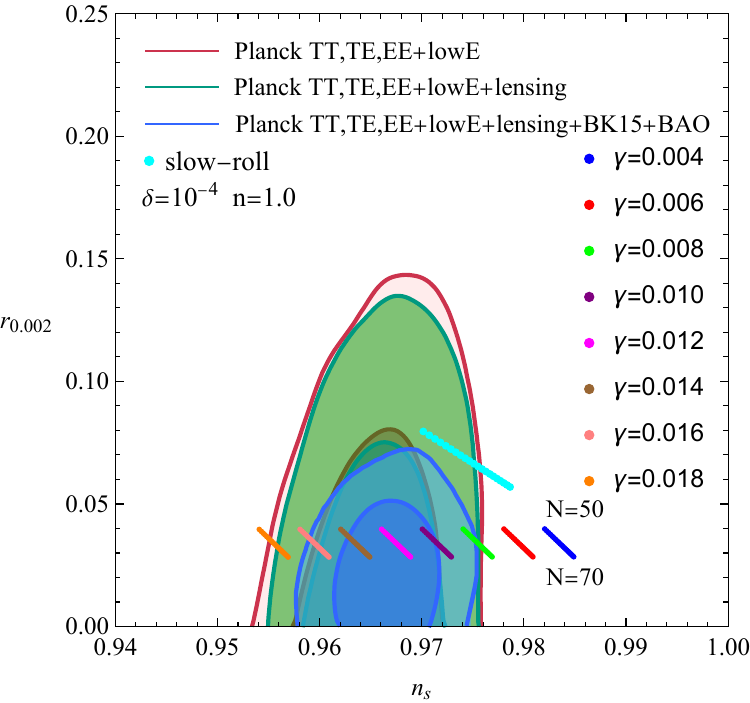}
\includegraphics[width=0.45\textwidth]{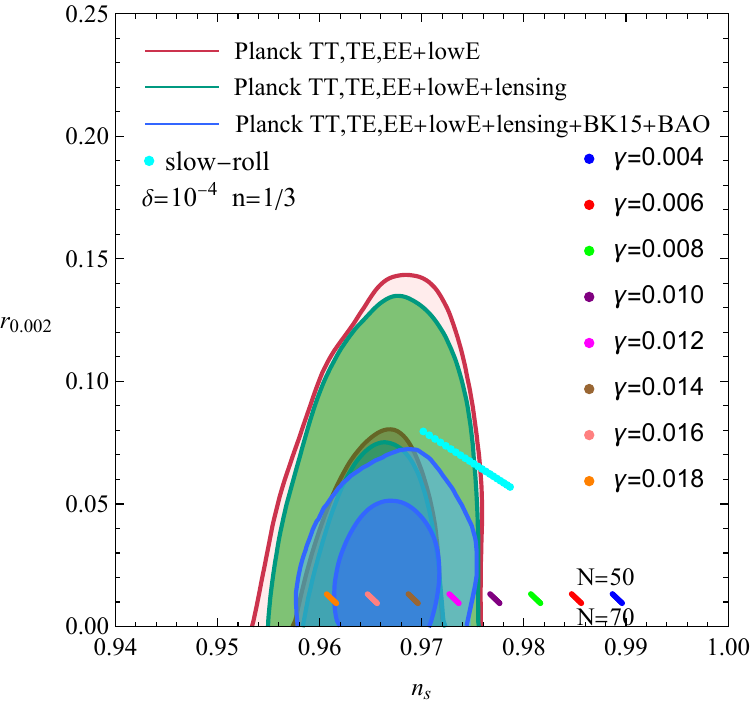}
\includegraphics[width=0.45\textwidth]{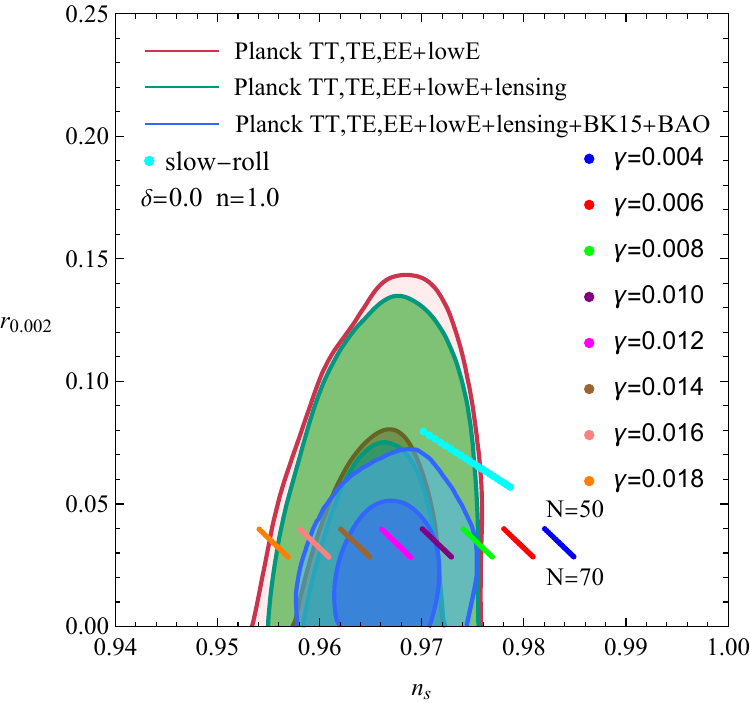}
\includegraphics[width=0.45\textwidth]{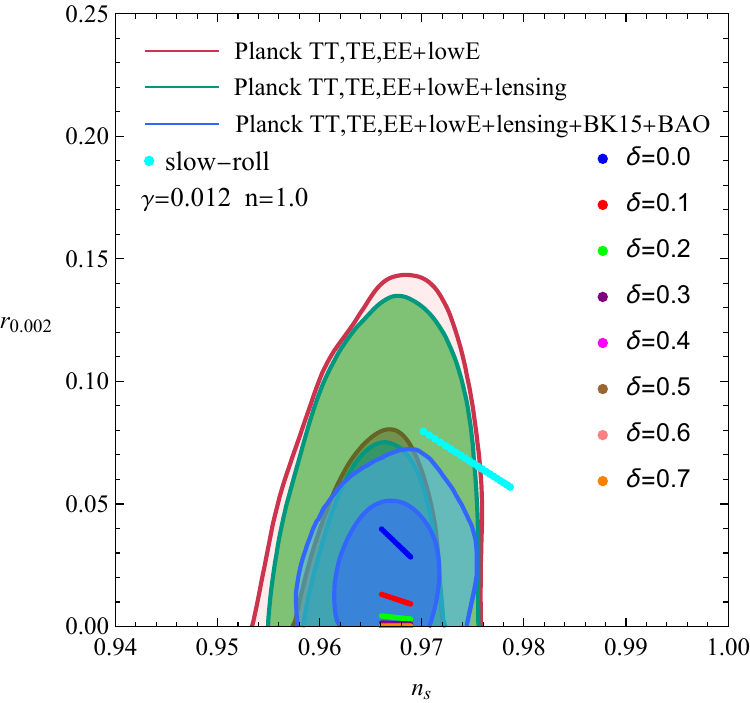}
\caption{\label{Fig2} Predictions of the power-law potential ~(\ref{Vp}) in $r_{0.002}-n_{s}$ plane.}
\end{center}
\end{figure*}

With fixed value of $n$, we depict the predictions of the power-law potential ~(\ref{Vp}) in $r_{0.002}-n_{s}$ plane in Fig.~(\ref{Fig2}), where we overlap our analytical results with Planck 2018 data, while the cyan dots denote the case in standard slow-roll inflation~\cite{Planck2020}. These figures show that as $N$ increases from $50$ to $70$, the value of $r$ decreases, and an increasing $\gamma$ leads to a smaller value of $n_{s}$. The case of $\delta=10^{-4}$ is indistinguishable from $\delta=0$, which corresponds to constant-roll inflation in general relativity. An increasing value of $\delta$ has a negligible effect on $n_{s}$ but causes a significant decrease in $r$. With this comparison with the observation, the results show a good consistency for a specific range of the constant-roll parameter $\gamma$ with Planck 2018 data. It is evident that an increased value of $n$ will lead to mismatch with the observations, and the value of $\gamma$ is determined by $n$ and increases as $n$ decreases. It is obvious that the constant-roll parameter $\gamma$ and the model parameter $\delta$ have a significant influence on $(r_{0.002},n_{s})$ behavior, exhibiting pronounced deviations from slow-roll inflation predictions. For the case where $\delta=10^{-4}$ and $n=1$, $\gamma$ takes the value $0.012$. For the case where $\gamma=0.012$ and $n=1$, the range $0 \leq \delta \leq 0.7$ can be supported by Planck 2018 data. It is interesting to note that $n=1$ is also obtained in constant-roll inflation with Tsallis holographic dark energy~\cite{Keskin2023}, whereas the power-law potential in slow-roll inflation within general relativity has been excluded by Planck 2018 data~\cite{Planck2020}, as shown by the cyan dots.

\subsection{Periodic potential}

\begin{figure*}[htp]
\begin{center}
\includegraphics[width=0.45\textwidth]{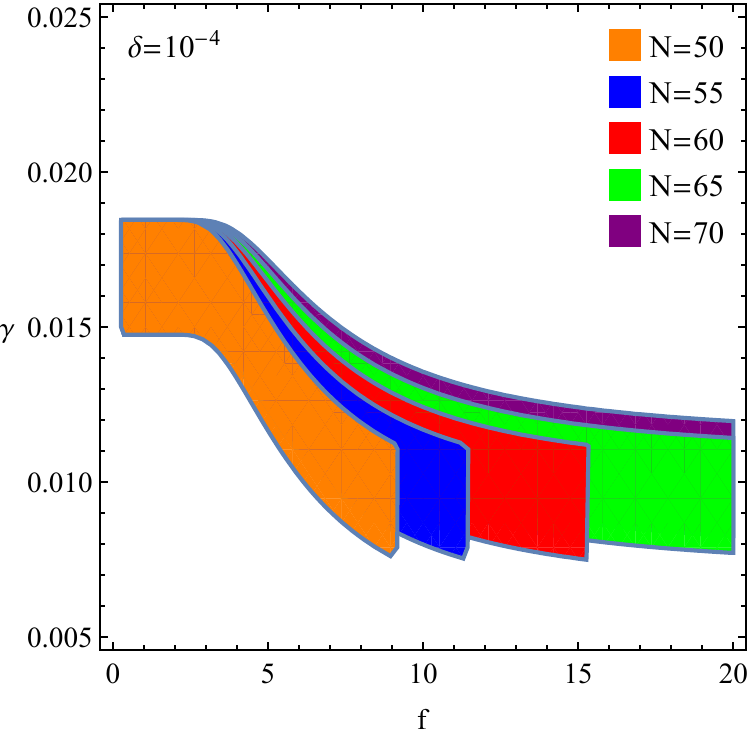}
\includegraphics[width=0.45\textwidth]{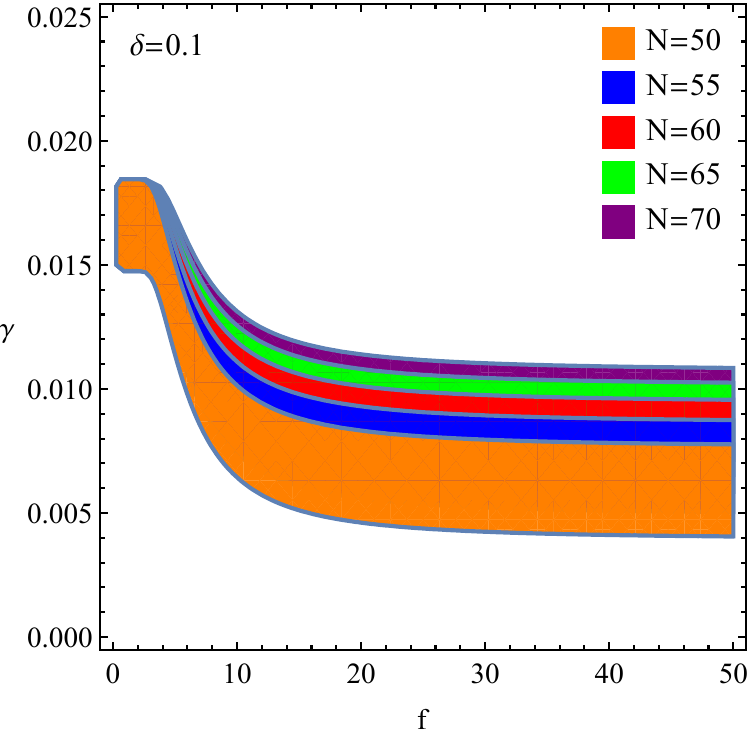}
\caption{\label{Fig3} Prediction regions of the periodic potential ~(\ref{Vp2}) in $\gamma-f$ plane.}
\end{center}
\end{figure*}

The next potential we examine is the periodic potential
\beq\label{Vp2}
V=V_{0}\Big[ 1+\cos\Big(\frac{\phi}{f}\Big) \Big],
\eeq
where $V_{0}$ and $f$ are positive parameters with $f$ controlling the curvature of the potential. This potential, known as the natural inflation potential~\cite{Freese1990, Adams1993}, derives from string theory frameworks and emerges naturally when the inflaton field corresponds to a pseudo-Nambu-Goldstone boson, i.e., an axion~\cite{Freese1990}. It has been studied in contexts including slow-roll inflation~\cite{Bassett2006, Planck2020, Tahmasebzadeh2016}, reheating~\cite{ Cook2015}, and PBH formation~\cite{Gao2021, Cook2023}. Using the same analytical method as that in the previous subsection, we obtain the first constant-roll parameter $\epsilon_{1}$ expressed as
\beq
\epsilon_{1} \approx \frac{1}{2(2-\delta)} \frac{3^{\frac{\delta-4}{\delta-2}}}{(3+\gamma)^{2} f^{2}} \frac{1}{e^{\frac{3N}{(3+\gamma) f^{2}}}\big[ 1+\frac{9}{4(3+\gamma)^{2} f^{2}} \big]-1}.
\eeq
Then, we plot the prediction regions in $\gamma-f$ plane with different e-folds number $N$, which is shown in Fig.~(\ref{Fig3}) with $\delta=10^{-4}$ and $\delta=0.1$. These figures demonstrate that $\gamma$ takes a small value, and $f$ is determined by the e-folds number $N$. With the increase of $f$, a large $N$ is required. A larger value of $\delta$ expands the range of the $\gamma-f$ plane. For $\delta = 10^{-4}$, $f<9.17$ is required, while $f<27.99$ is required for $\delta = 0.1$.

\begin{figure*}[htp]
\begin{center}
\includegraphics[width=0.45\textwidth]{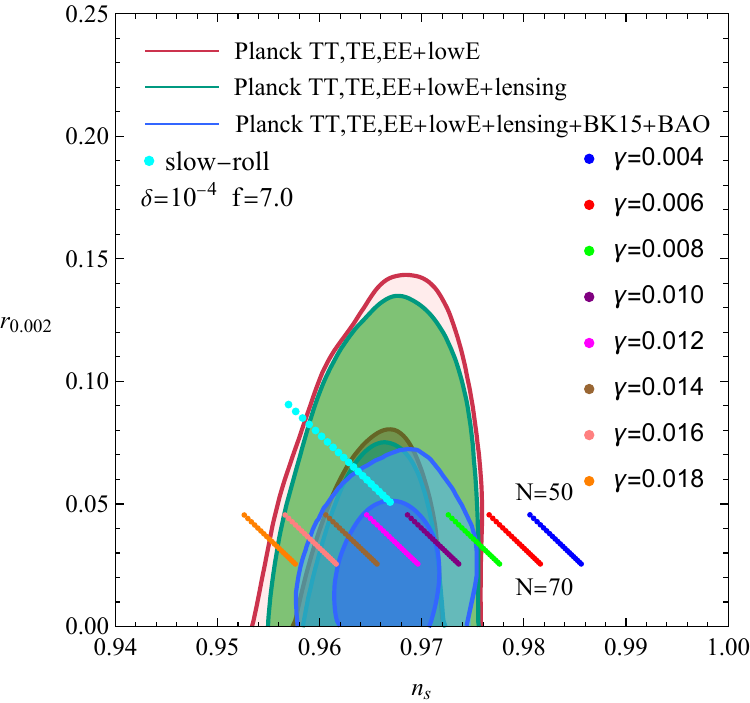}
\includegraphics[width=0.45\textwidth]{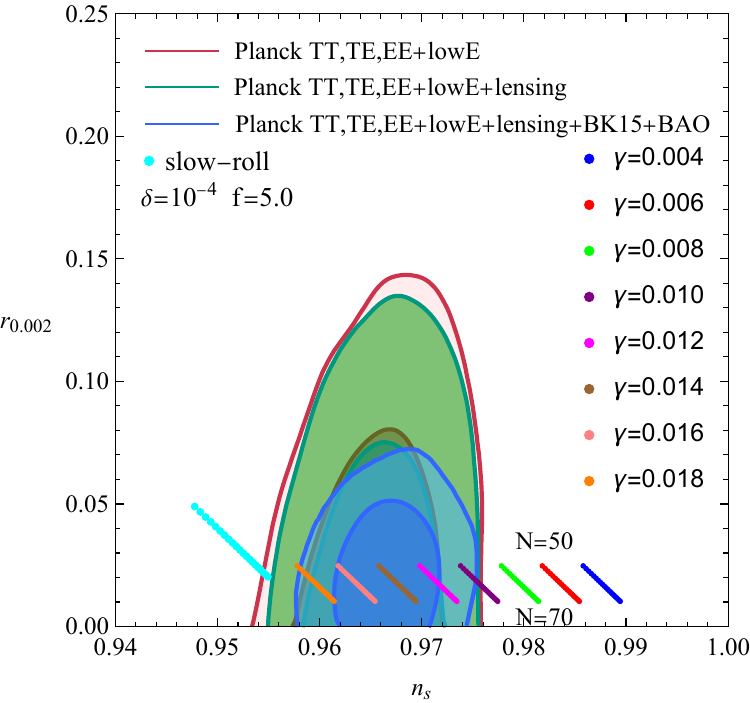}
\includegraphics[width=0.45\textwidth]{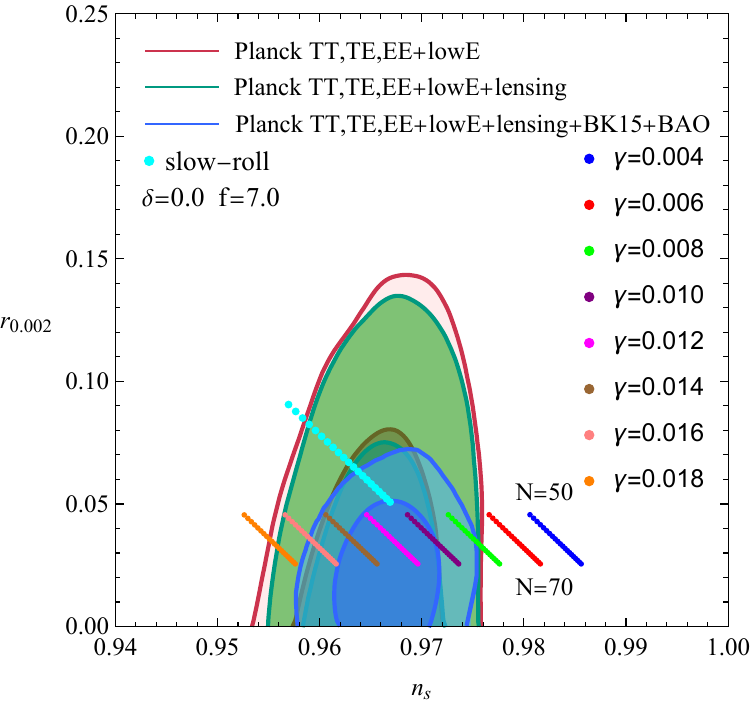}
\includegraphics[width=0.45\textwidth]{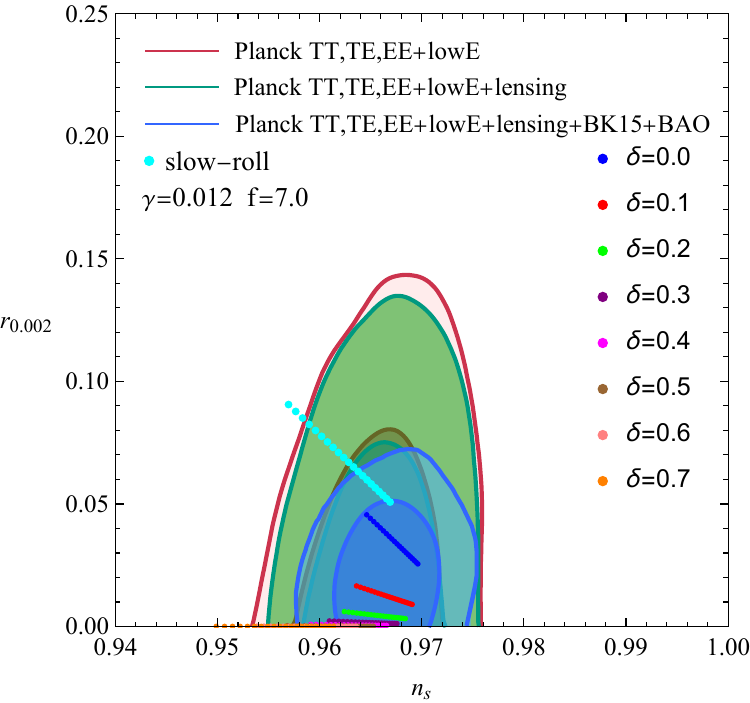}
\caption{\label{Fig4} Predictions of the periodic potential ~(\ref{Vp2}) in $r_{0.002}-n_{s}$ plane.}
\end{center}
\end{figure*}

Then, fixing the value of $f$, we plot the predictions of the periodic potential ~(\ref{Vp2}) in $r_{0.002}-n_{s}$ plane in Fig.~(\ref{Fig4}), where we overlap our analytical results with Planck 2018 data, and the cyan dots denote the case in standard slow-roll inflation~\cite{Planck2020}. Similar to the power-law potential, these figures also show that the value of $r$ decreases as $N$ varies from $50$ to $70$, and an increasing $\gamma$ leads to a smaller value of $n_{s}$. The case of $\delta=10^{-4}$ cannot be distinguished from $\delta=0$, which corresponds to constant-roll inflation in general relativity. An increasing value of $\delta$ leads to a decrease in $n_{s}$ and a significant decrease in $r$. By comparing with Planck 2018 data, the results show a good consistency for a specific range of the constant-roll parameter $\gamma$. Both the constant-roll parameter $\gamma$ and the model parameter $\delta$ have a significant influence on $(r_{0.002},n_{s})$ behavior, exhibiting significant deviations from slow-roll inflation predictions. For the case where $\delta=10^{-4}$ and $f=7$, $\gamma$ takes the value $0.012$. For the case where $\gamma=0.012$ and $f=7$, the range $0 \leq \delta \leq 0.2$ is supported by Planck 2018 data. These figures indicate that increasing $f$ leads to mismatch with observations, and $\gamma$ increases as $f$ decreases. Additionally, the periodic potential in slow-roll inflation within general relativity is disfavored by Planck 2018 data~\cite{Planck2020}, as shown by the cyan dots.

\subsection{Hilltop potential}

The last potential we analyze is the hilltop potential, which plays a significant role in inflationary cosmology and takes the form
\beq\label{Vp3}
V=V_{0}\Big[1-\Big(\frac{\phi}{\mu}\Big)^{p}\Big],
\eeq
as defined in hilltop inflation~\cite{Boubekeur2005}. Here, $V_{0}$, $\mu$, and $p$ are constant parameters of the model. This potential may arise in spontaneous symmetry breaking~\cite{Boubekeur2005, Bassett2006} and shares applications in slow-roll inflation analysis~\cite{Bassett2006, Planck2020, Tahmasebzadeh2016}, alongside reheating mechanisms~\cite{Bassett2006, Cook2015, Goswami2018}, and PBH studies~\cite{Cook2023}. Planck 2018 reported that the quartic potential provides a better fit than the quadratic potential, and $\mu>10$ is required in the quartic case~\cite{Planck2020}. To analyze the constant-roll inflation in the hilltop potential, we adopt the same analytical method as that in the previous subsections. However, due to the unique characteristics of hilltop potential and the computational complexity, we cannot obtain an analytical solution and need to employ numerical methods for complete analysis.

\begin{figure*}[htp]
\begin{center}
\includegraphics[width=0.45\textwidth]{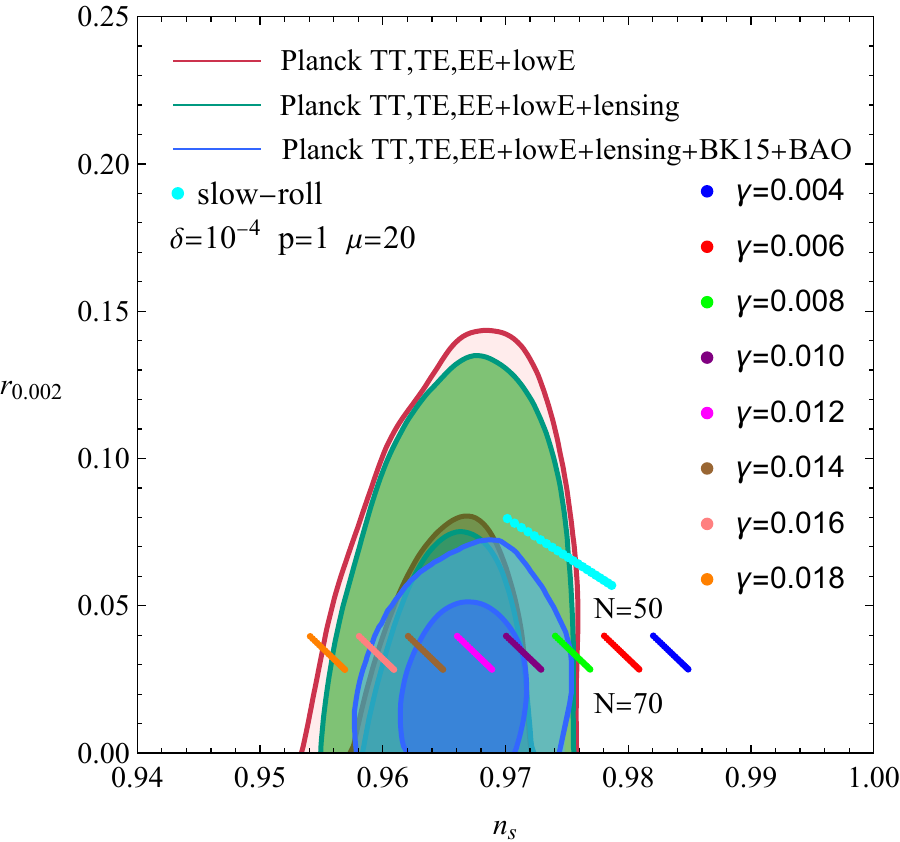}
\includegraphics[width=0.45\textwidth]{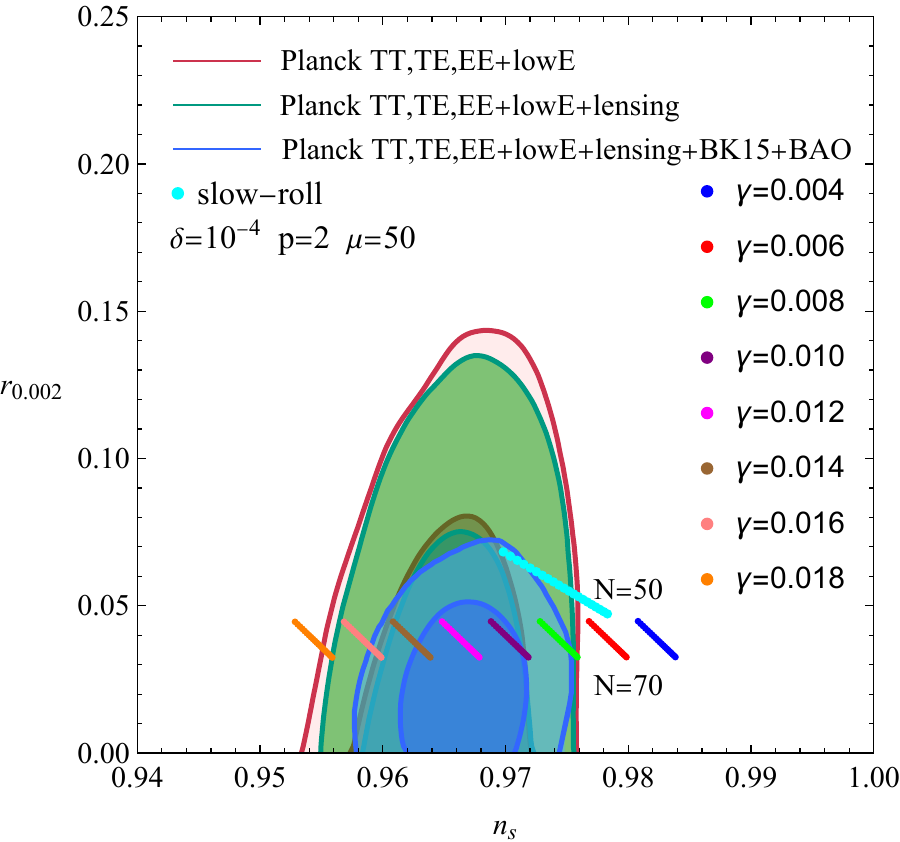}
\includegraphics[width=0.45\textwidth]{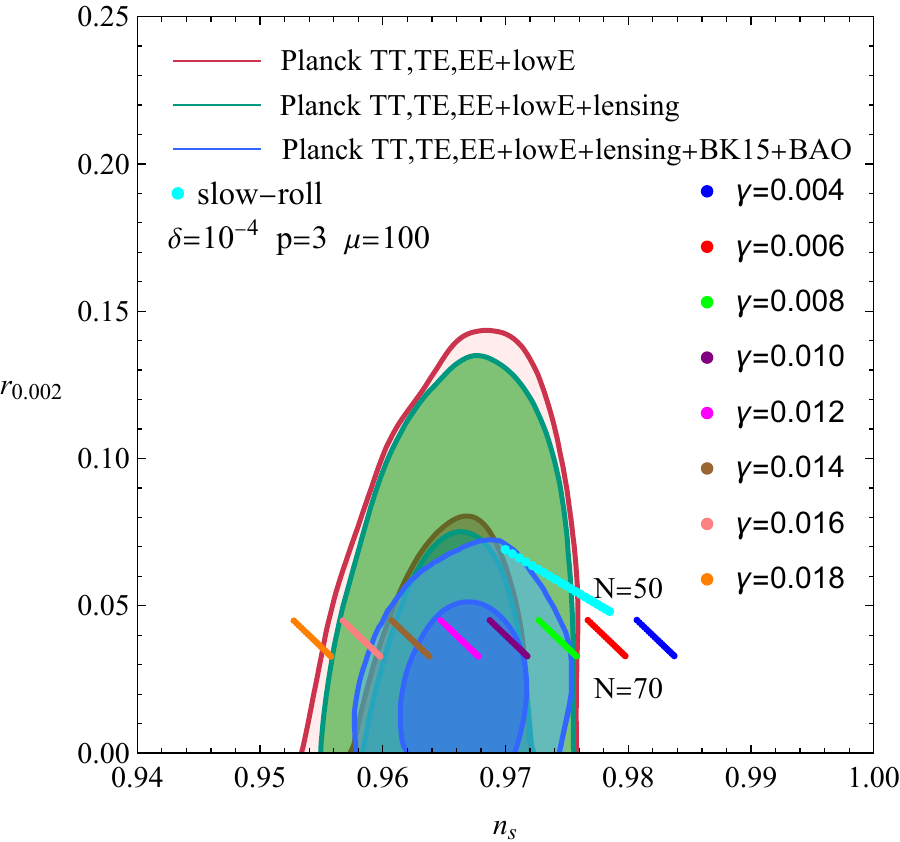}
\includegraphics[width=0.45\textwidth]{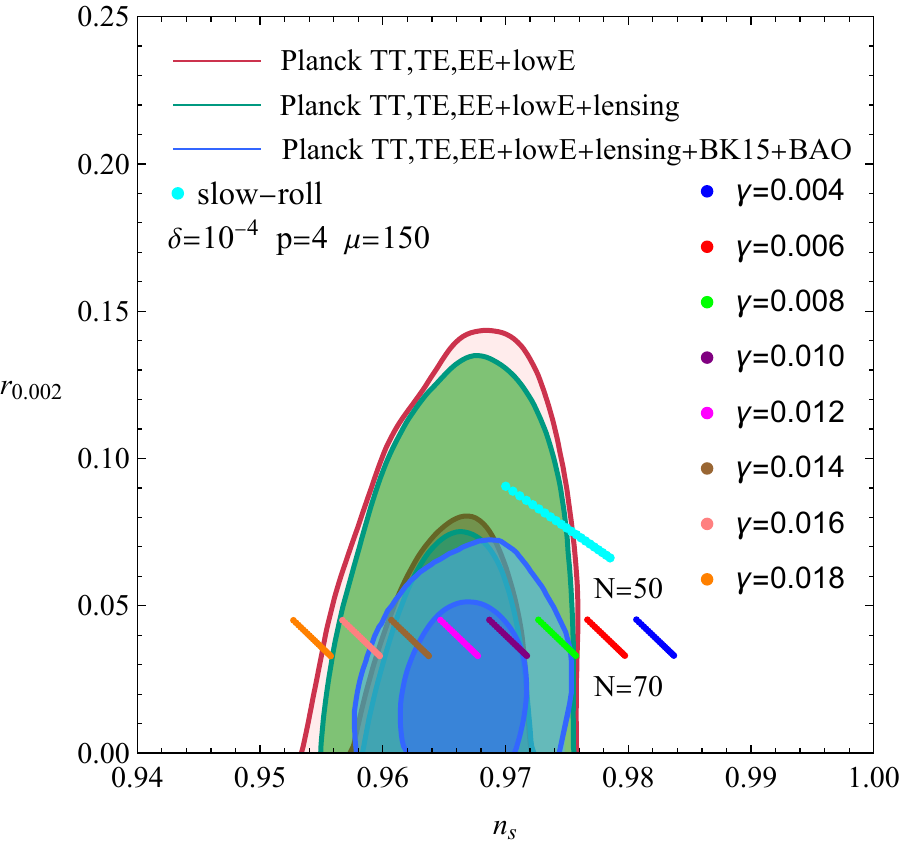}
\caption{\label{Fig5} Predictions of the hilltop potential ~(\ref{Vp3}) in $r_{0.002}-n_{s}$ plane with $\delta=10^{-4}$.}
\end{center}
\end{figure*}

\begin{figure*}[htp]
\begin{center}
\includegraphics[width=0.45\textwidth]{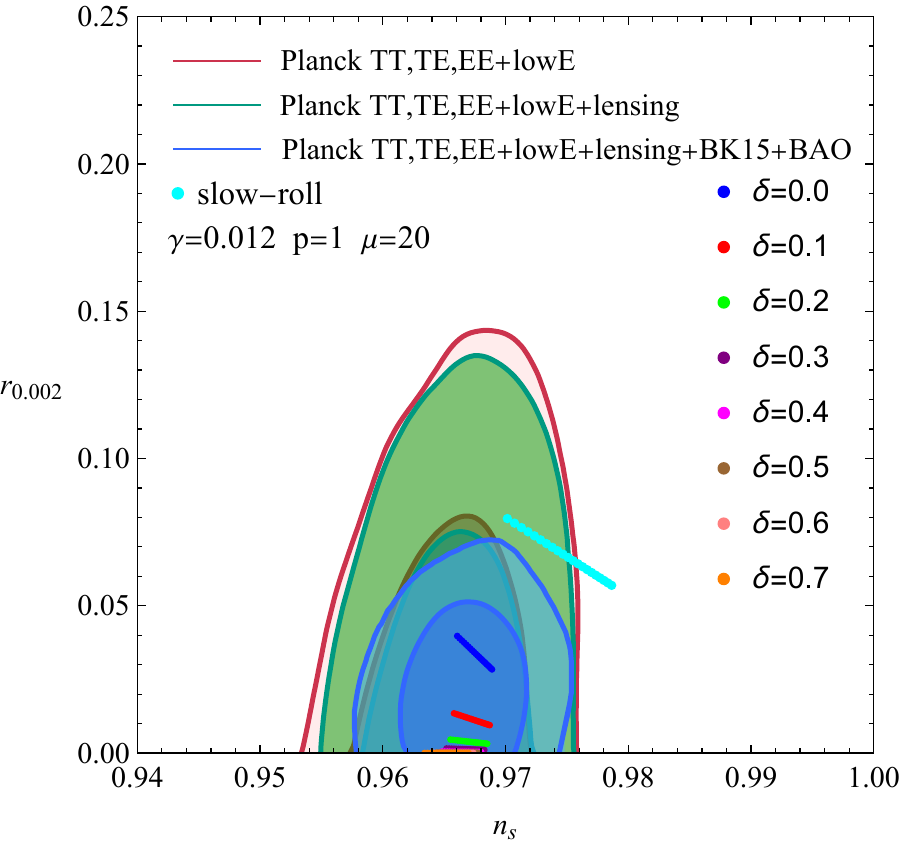}
\includegraphics[width=0.45\textwidth]{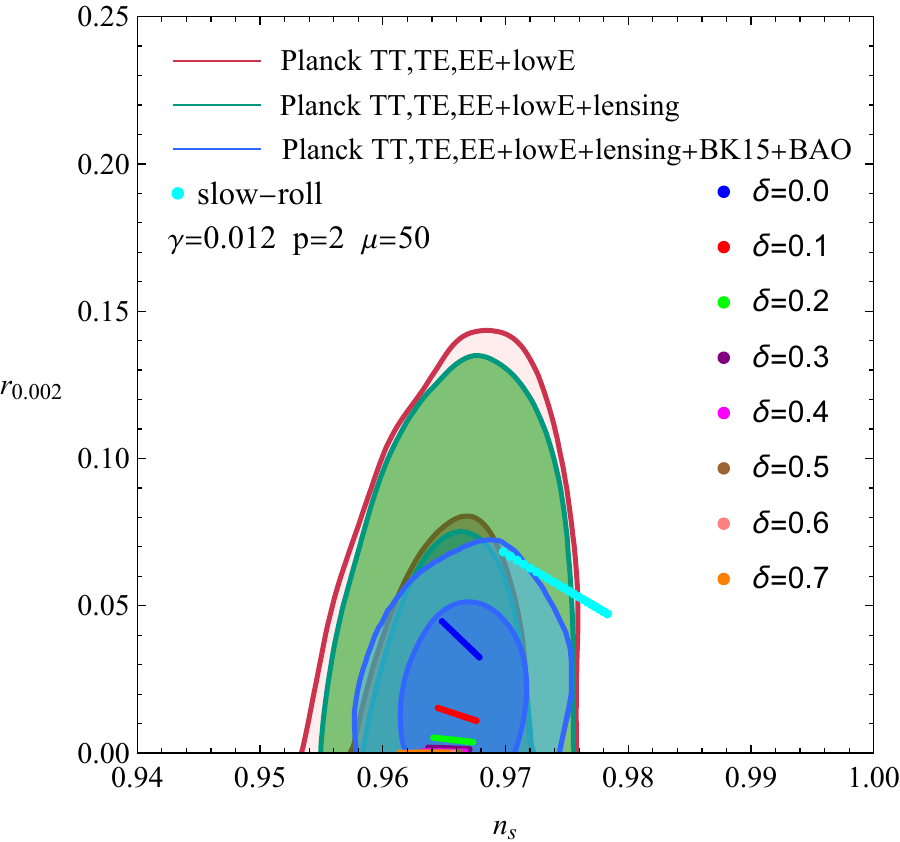}
\includegraphics[width=0.45\textwidth]{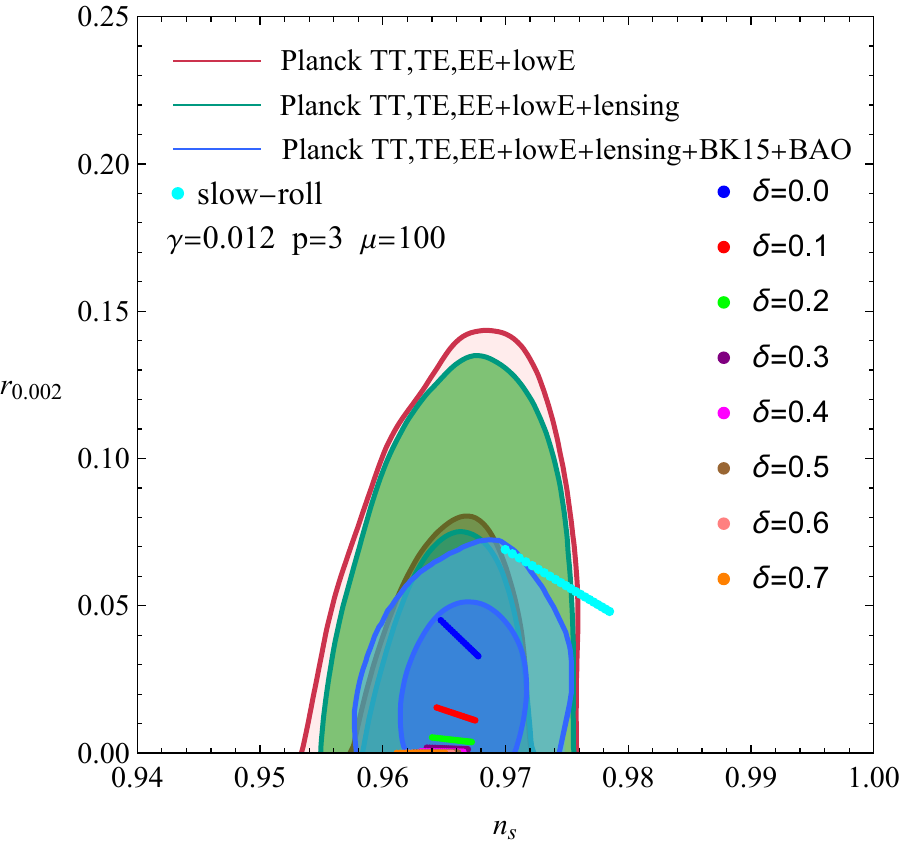}
\includegraphics[width=0.45\textwidth]{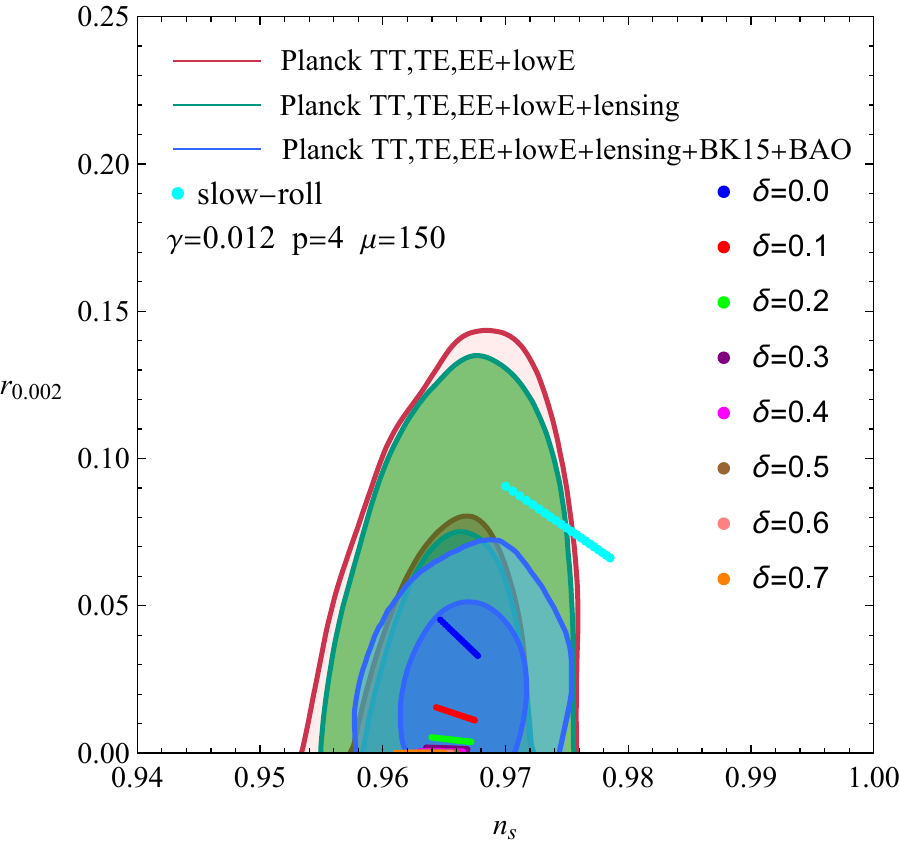}
\caption{\label{Fig5e} Predictions of the hilltop potential ~(\ref{Vp3}) in $r_{0.002}-n_{s}$ plane with different values of $\delta$.}
\end{center}
\end{figure*}

Fixing the values of different parameters, we depict the predictions of the hilltop potential ~(\ref{Vp3}) in $r_{0.002}-n_{s}$ plane in Figs.~(\ref{Fig5}) and ~(\ref{Fig5e}) for some special cases, overlapping our analytical results with Planck 2018 data, where the cyan dots represent the case in standard slow-roll inflation~\cite{Planck2020}. These figures also show that the value of $r$ decreases as $N$ varies from $50$ to $70$, and an increasing $\gamma$ leads to a smaller value of $n_{s}$. To satisfy the observational constraints from Planck 2018 data, $\mu$ must increase with the increase of $p$. As shown in these figures, $p=1$ requires $\mu=20$ while $p=4$ requires $\mu=150$ to satisfy these constraints. Similar to the previous model, the case of $\delta=10^{-4}$ cannot be distinguished from $\delta=0$, and an increasing value of $\delta$ leads to a decrease in $n_{s}$ and a significant decrease in $r$. It is evident that the constant-roll parameter $\gamma$ and the model parameter $\delta$ directly affect the $(r_{0.002},n_{s})$ observational predictions, showing measurable discrepancies compared to the standard slow-roll inflation predictions. For all cases $\delta=10^{-4}$ with $p=1,2,3,4$, a consistent value of $\gamma=0.012$ is needed. For the case where $\gamma=0.012$, $p=1$ and $\mu=20$, the range $0 \leq \delta \leq 0.7$ is favored by Planck 2018 data; while for $\gamma=0.012$, $p=4$ and $\mu=150$, the range supported by Planck 2018 data becomes $0 \leq \delta \leq 0.5$. For the standard slow-roll inflation with the hilltop potential~\cite{Planck2020}, the cases in Figs.~(\ref{Fig5}) and ~(\ref{Fig5e}) are disfavored by Planck 2018 data, as shown by the cyan dots, while they can be supported in constant-roll inflation.

\subsection{Attractor behavior}

It is well established that inflationary models with canonical scalar fields exhibit cosmological attractor behavior~\cite{Remmen2013}. This behavior serves as a critical feature in viable scenarios by ensuring theoretical consistency across different initial conditions. The attractor behavior of the solution can be investigated through analytical and numerical methods. Using the analytical method, a stable attractor for constant-roll inflation in holographic dark energy is found~\cite{Mohammadi2022}. In this subsection, we analyze the attractor behavior of constant-roll inflation in the context of the Barrow entropy model and check whether our analytical solution is an attractor solution.

Combining the Klein-Gordon equation ~(\ref{s2}), the constant-roll condition ~(\ref{csc}), and the Friedmann equation ~(\ref{wr1})  with the power-law potential ~(\ref{Vp}), the periodic potential ~(\ref{Vp2}), and the hilltop potential ~(\ref{Vp3}) respectively, we obtain analytical solutions for constant-roll inflation in the context of the Barrow entropy model, which are shown as follows
\bea
&& \dot{\phi} = -\Big(\frac{8\pi G_{eff}}{3}\Big)^{-\frac{1}{2-\delta}} \frac{n}{\gamma+3} V^{\frac{1-\delta}{2-\delta}}_{0} \phi^{\frac{n(1-\delta)-(2-\delta)}{2-\delta}},\label{as1}\\
&& \dot{\phi} = \Big(\frac{8\pi G_{eff}}{3}\Big)^{-\frac{1}{2-\delta}} \frac{V_{0}}{(\gamma+3)f} \frac{\sin\big(\frac{\phi}{f}\big)}{\Big[V_{0}\big(1+\cos\big(\frac{\phi}{f}\big)\big)\Big]^{\frac{1}{2-\delta}}},\label{as2}\\
&& \dot{\phi} = \Big(\frac{8\pi G_{eff}}{3}\Big)^{-\frac{1}{2-\delta}} \frac{V_{0} p}{(\gamma+3)\mu} \frac{\big(\frac{\phi}{\mu}\big)^{p-1}}{\Big[V_{0}\big(1-\big(\frac{\phi}{\mu}\big)^{p}\big)\Big]^{\frac{1}{2-\delta}}}.\label{as3}
\eea

\begin{figure*}[htp]
\begin{center}
\includegraphics[width=0.325\textwidth]{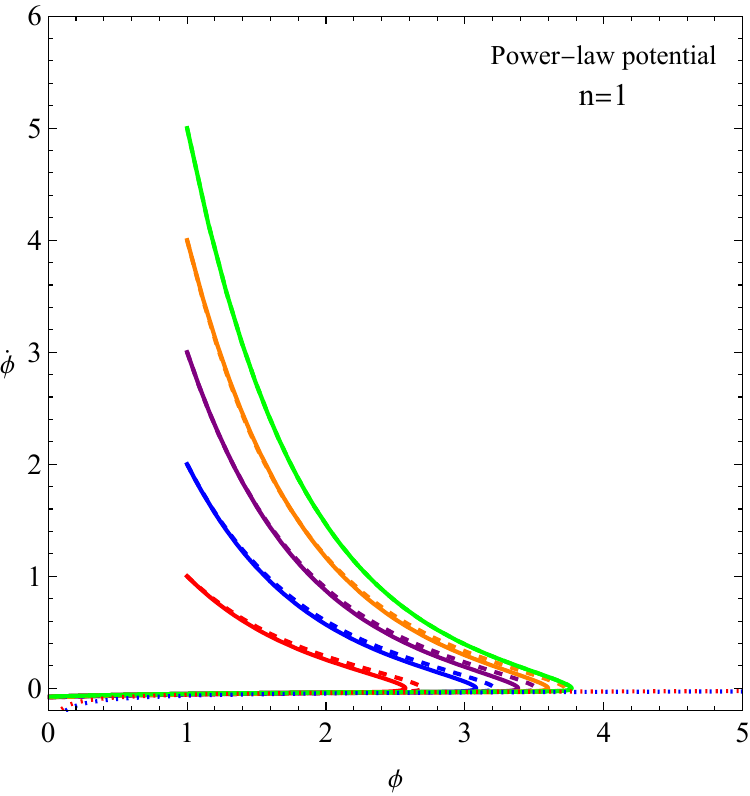}
\includegraphics[width=0.325\textwidth]{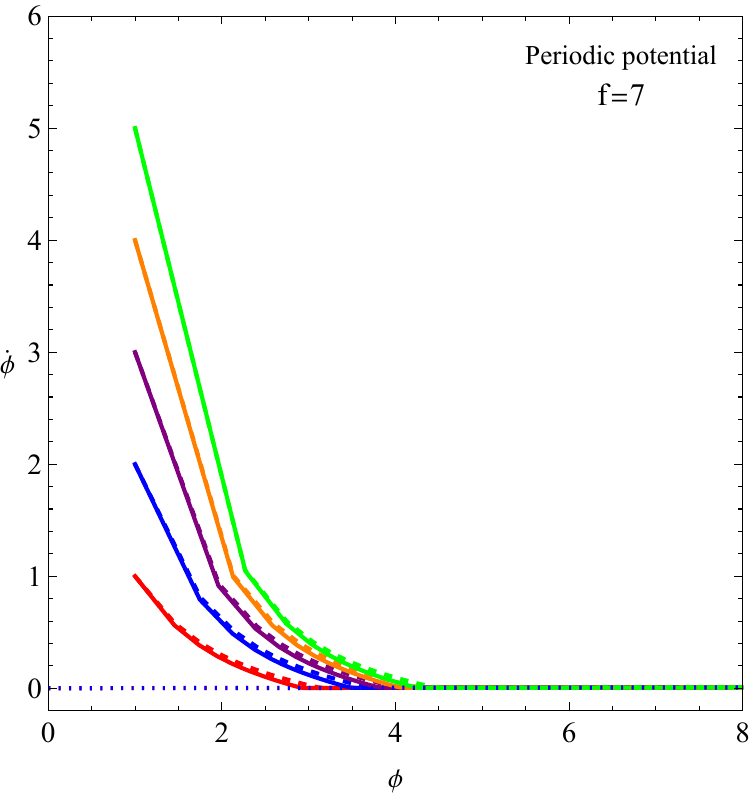}
\includegraphics[width=0.325\textwidth]{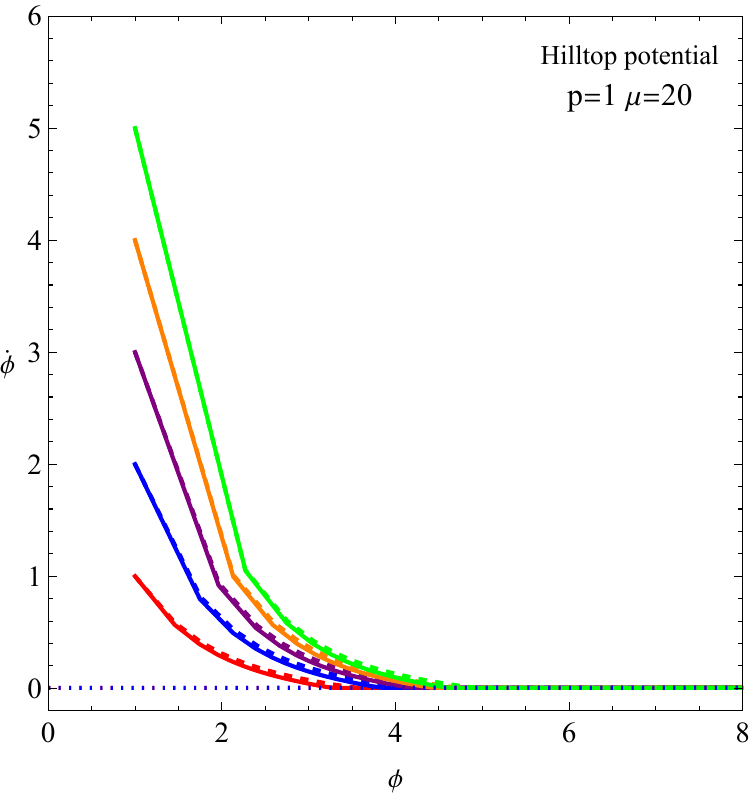}
\caption{\label{Fig50} Phase space diagram for the power-law potential, the periodic potential, and the hilltop potential. Solid and dashed lines denote the cases $\delta=10^{-4}$ and $\delta=0.1$, while red and blue dotted lines show the analytical solutions for $\delta=10^{-4}$ and $\delta=0.1$.}
\end{center}
\end{figure*}

Then, by numerically solving the Klein-Gordon equation ~(\ref{s2}) and the Friedmann equation ~(\ref{H1}) for the power-law potential ~(\ref{Vp}), the periodic potential ~(\ref{Vp2}), and the hilltop potential ~(\ref{Vp3}) under different initial conditions respectively, we construct the phase space diagram as shown in Fig.~(\ref{Fig50}), where solid and dashed lines denote $\delta=10^{-4}$ and $\delta=0.1$, with red and blue dotted lines representing the analytical solutions for $\delta=10^{-4}$ and $\delta=0.1$ respectively. These figures show that regardless of the initial conditions for constant-roll inflation, the kinetic energy rapidly decays to a tiny value, driving the inflaton field into a potential dominated constant-roll state. The evolutionary trajectories exhibit strong convergence, eventually merging toward the analytical solution marked by the red and blue dotted line. Thus, in these models, the analytical solution is an attractor solution; the evolutionary trajectories converge to it.

\section{Primordial black holes}

In the previous section, we have studied the constant-roll inflation within the Barrow entropy model with power-law potential, periodic potential, and hilltop potential and found that they can be supported by the Planck 2018 data. During the radiation-dominated era after inflation, the enhanced primordial curvature perturbations at small scales may form PBHs through gravitational collapse when they re-enter the horizon. In standard slow-roll inflation, the PBHs formation probabilities remain exponentially suppressed because the curvature perturbation power spectrum fails to reach critical amplitude thresholds. In this section, we will discuss the formation and evolution of PBHs within the Barrow entropy model.

\subsection{Formation}

The mass of PBHs is related to the horizon mass when curvature perturbations with comoving wavenumber $k$ reenter the cosmological horizon~\cite{Motohashi2017p}
\beq\label{Mk}
M(k) \simeq M_{\odot}\Big(\frac{\gamma}{0.2}\Big)\Big(\frac{g_{*}}{10.75}\Big)^{-\frac{1}{6}}\Big(\frac{k}{1.9 \times 10^{6} Mpc^{-1}}\Big)^{-2},
\eeq
where $M_{\odot}$ represents the Solar mass, $\gamma \simeq (\frac{1}{\sqrt{3}})^{3}$ is the ratio of PBH mass to horizon mass and indicates the efficiency of collapse, and $g_{*}=106.75$ corresponds to the effective degrees of freedom in the energy density during the PBH formation. The production rate of PBHs with mass $M(k)$ is given by~\cite{Young2014, Tada2019}
\beq\label{betaM}
\beta(M)=\int_{\delta_{c}}\frac{d\delta}{\sqrt{2\pi \sigma^{2}(M)}}e^{-\frac{\delta^{2}}{2\sigma^{2}(M)}}=\frac{1}{2}\text{erfc}\Big(\frac{\delta_{c}}{\sqrt{2\sigma^{2}(M)}}\Big),
\eeq 
where $\text{erfc}$ represents the complementary error function, $\delta_{c} \simeq 0.4$ is the threshold of the density perturbations for the PBH formation~\cite{Musco2013, Harada2014}, and $\sigma^{2}(M)$ denotes the variance of the coarse-grained density contrast, smoothed over the scale k, defined as~\cite{Young2014}
\beq\label{sigma2}
\sigma^{2}(M(k))=\frac{16}{81}\int \frac{dq}{q} W^{2}(q/k)(q/k)^{4}\mathcal{P}_{\mathcal{R}}(q),
\eeq
where $W(x)=e^{-\frac{x^{2}}{2}}$ is the Gaussian window function, and the power spectrum of the primordial curvature perturbations $\mathcal{P}_{\mathcal{R}}(k)$ can be solved from the Mukhanov-Sasaki equation~(\ref{MSE}). The current fractional energy density of PBHs with respect to the total dark matter density is given as~\cite{Carr2016}
\beq\label{OPOD}
\frac{\Omega_{PBH}}{\Omega_{DM}}=\int\frac{dM}{M}f(M)
\eeq
with
\beq\label{fM}
f(M)=\frac{1}{\Omega_{DM}}\frac{d\Omega_{PBH}}{d\ln M} \simeq \frac{\beta(M)}{1.84 \times 10^{-8}}\Big(\frac{\gamma}{0.2}\Big)^{\frac{3}{2}} \Big(\frac{10.75}{g_{*}}\Big)^{\frac{1}{4}}\Big(\frac{0.12}{\Omega_{DM}h^{2}}\Big)\Big(\frac{M}{M_{\odot}}\Big)^{-\frac{1}{2}}.
\eeq
Here, the value of $\Omega_{DM}h^{2}$ is given by Planck 2018 data~\cite{Planck2020a}. Based on the above equations, one can calculate PBH abundances and note that producing a sizable amount of PBHs requires the typical primordial curvature perturbations to be significant, $\mathcal{P_{R}} \sim O(10^{-2})$ on small scales. This value is about $7$ orders of magnitude larger than the corresponding perturbations $\mathcal{P_{R}} \sim O(10^{-9})$ observed on CMB scales. To obtain enhanced primordial curvature perturbations at small scales, we consider that there is a small part with a periodic structure in the inflation potentials in the previous section, and the power-law potential ~(\ref{Vp}), periodic potential ~(\ref{Vp2}), and hilltop potential ~(\ref{Vp3}) take the following form
\bea
&& V=V_{0}\phi^{n}+\delta V,\\
&& V=V_{0}\Big[ 1+\cos\Big(\frac{\phi}{f}\Big) \Big]+\delta V,\\
&& V=V_{0}\Big[1-\Big(\frac{\phi}{\mu}\Big)^{p}\Big]+\delta V,
\eea
with
\beq
\delta V=\xi\sin\Big(\frac{\phi}{\phi_{*}}\Big)[1+\tanh(10^{6}(\phi_{e}-\phi)(\phi-\phi_{s}))],
\eeq
where $\delta V$ represents a small periodic-structure contribution to the potential ($\delta V \ll V$), with $\xi$ quantifying the magnitude of the structure, $\phi_{*}$ governing its characteristic period, and $\phi_{s}$ and $\phi_{e}$ specifying the starting and ending points of the structural feature respectively. This small periodic structure in the inflationary potential may amplify primordial curvature perturbations at small scales through parametric resonance mechanisms~\cite{Cai2020}. The parametric resonance can be achieved either by adding a periodic correction to the inflationary potential~\cite{Cai2020} or by introducing a periodic sound speed~\cite{Cai2018}. Since its proposal, this mechanism has been widely applied to enhance both primordial curvature perturbations and gravitational wave amplitudes~\cite{Zhou2020, Peng2021, Cai2021, Cai2024, Cai2019, Chen2019, Chen2020, Addazi2022, Yu2024}. The amplified primordial curvature perturbations could subsequently lead to observationally significant PBH populations upon horizon re-entry during the radiation-dominated era.

\begin{figure*}[htp]
\begin{center}
\includegraphics[width=0.46\textwidth]{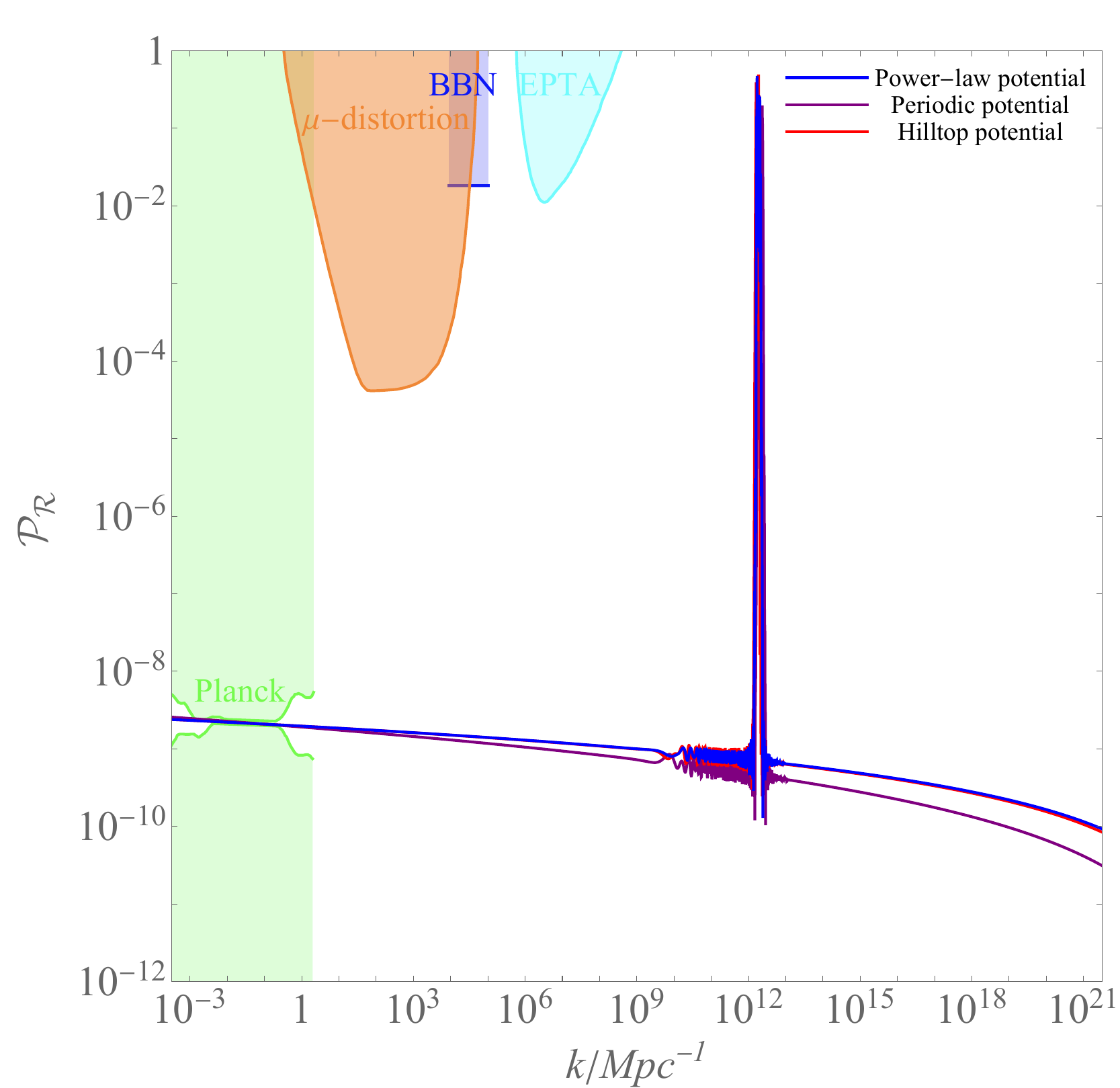}
\includegraphics[width=0.44\textwidth]{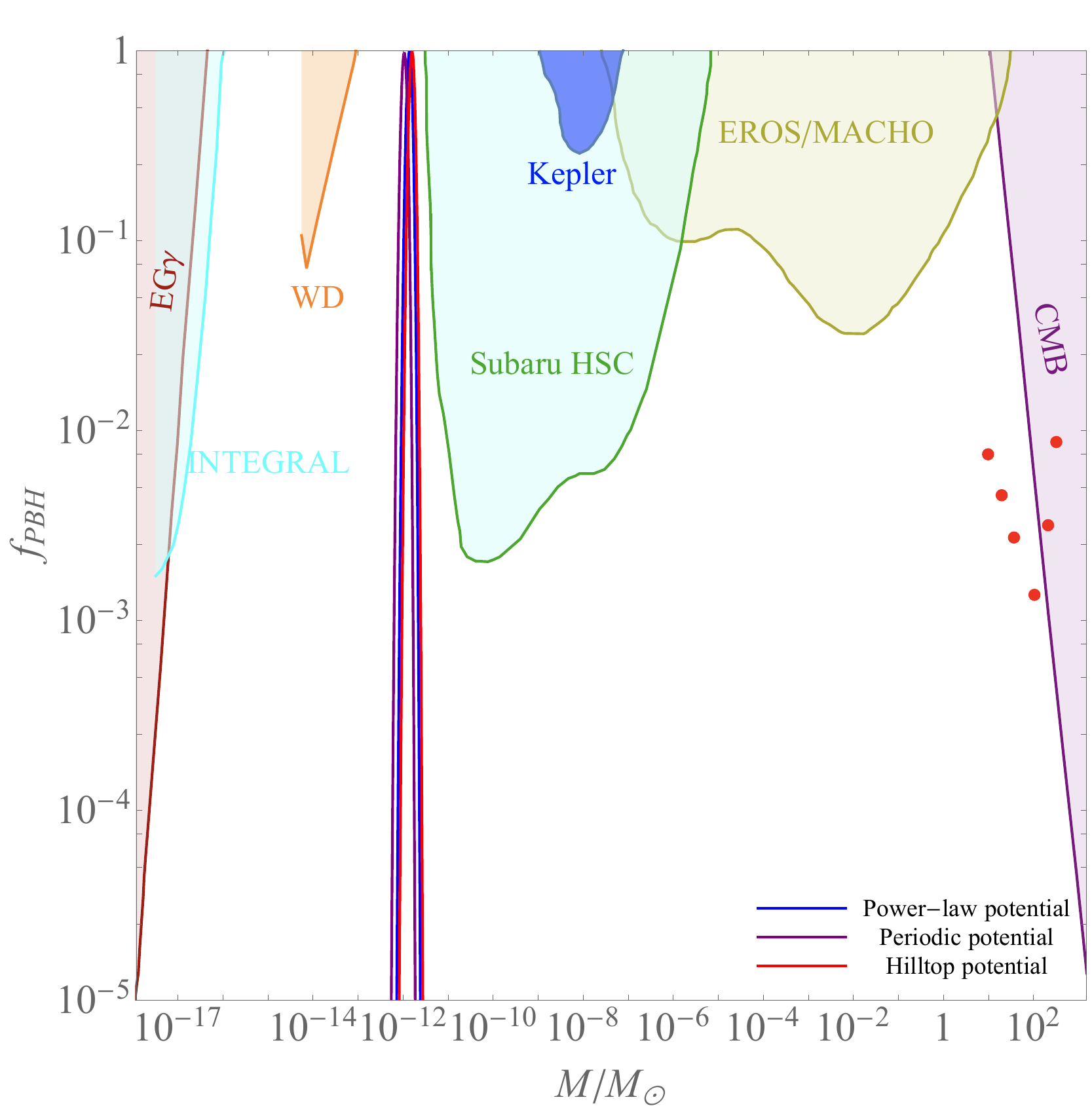}
\caption{\label{Fig6} Results for the primordial curvature perturbation power spectrum $\mathcal{P_{R}}$ and PBHs abundances $f_{PBH}$.}
\end{center}
\end{figure*}

Choosing the parameter of Barrow entropy model $\delta=10^{-4}$ and the potential parameters $n=1$, $f=7$, $p=1$, and $\mu=20$ according to the results in the previous section, fixing the energy scale $V_{0}=0.92 \times 10^{-10}$ and the period parameter $\phi_{*}=10^{-4}$, and adopting the values of the structural parameters $\xi$, $\phi_{s}$, and $\phi_{e}$ listed in Tab.~(\ref{Tab1}), we numerically solve the background equations~(\ref{H1}), ~(\ref{s2}), and ~(\ref{H2}) and the perturbation equation~(\ref{MSE}) in the flat universe under the constant-inflation condition~(\ref{csc}) to obtain the scalar power spectrum, as shown in the left panel of Fig.~(\ref{Fig6}). In this figure, we superimpose our numerical results on the observational data from CMB~\cite{Planck2020a}, $\mu$-distortion of CMB~\cite{Fixsen1996}, big-bang nucleosynthesis (BBN~\cite{Inomata2016}), and European Pulsar Timing Array (EPTA~\cite{Inomata2019}). For the power-law, periodic, and hilltop potential models, the numerical results of the e-folds number$N$, the scalar spectral index $n_{s}$, the tensor-to-scalar ratio $r$, the peak scales $k_{peak}$, and the peak values of the primordial curvature perturbation power spectrum $\mathcal{P^{\textit{peak}}_{R}}$ are shown in Tab.~(\ref{Tab1}). It can be seen that the numerical results of the primordial curvature perturbation power spectra not only satisfy the current observational constraints but also achieve a peak amplitude of the order of $10^{-2}$ at small scales. This enhancement indicates that these models can generate a sufficient number of PBHs, providing theoretical support for PBH formation.

\begin{table}
\caption{\label{Tab1} Model parameters and the numerical results.}
\centering
\resizebox{\textwidth}{!}{
\begin{tabular}{|c|c|c|c|c|c|c|c|c|c|c|c|c|}
  \hline
  \hline
  $Potential$ & $\xi(10^{-14})$ & $\phi_{e}$ & $\phi_{s}$ & $N$ & $n_{s}$ & $r$ & $k_{peak}(10^{12})$ & $\mathcal{P^{\textit{peak}}_{R}}(10^{-12})$ & $M_{peak}/M_{\odot}(10^{-12})$ & $f^{peak}_{PBH}$ & $\Omega_{PBH}/\Omega_{DM}$ & $f_{peak}(Hz)$\\
  \hline
  $Power-law$ & 35.367 & 8.30 &  8.35 & 59.99 & 0.9676 & 0.0335 & 1.704 & 0.473 & 1.40 & 0.979 & 0.334 & $3.0 \times 10^{-3}$\\
  \hline
  $Periodic$  & 5.450 & 10.79 & 10.84 & 59.97 & 0.9676 & 0.0336 & 1.917 & 0.262 & 1.10 & 0.982 & 0.337 & $3.6 \times 10^{-3}$\\
  \hline
  $Hilltop$   & 1.7792 & 11.70 & 11.75 & 59.98 & 0.9676 & 0.0338 & 1.857 & 0.493 & 1.60 & 0.981 & 0.331 & $3.1 \times 10^{-3}$\\
  \hline
  \hline
\end{tabular}}
\end{table}

Then, combining the numerical results of the primordial curvature perturbation power spectra in the left panel of Fig.~(\ref{Fig6}) with Eqs.~(\ref{Mk}), ~(\ref{betaM}), ~(\ref{sigma2}), ~(\ref{OPOD}), and ~(\ref{fM}), we obtain the peak masses of PBHs $M_{peak}$, the PBHs abundance $f^{peak}_{PBH}$, and the current fraction of PBHs in dark matter density $\Omega_{PBH}/\Omega_{DE}$, as shown in the right panel of Fig.~(\ref{Fig6}) and Tab.~(\ref{Tab1}). In the right panel of Fig.~(\ref{Fig6}), we superimpose our numerical results on observational data from the extragalactic gamma ray background (EG$\gamma$~\cite{Carr2010}), the white dwarfs explosion (WD~\cite{Graham2015}), the Galactic Center 511 keV gamma-ray line (INTEGRAL~\cite{Laha2019}), the microlensing events observed by the Kepler satellite (Kepler~\cite{Griest2013}), Subaru HSC~\cite{Niikura2019}, EROS/MACHO~\cite{Tisserand2007}, and CMB~\cite{Poulin2017}. The detailed numerical results for the power-law, periodic, and hilltop potential models in Tab.~(\ref{Tab1}) show that PBHs with a mass of approximately $10^{-12} M_{\odot}$ exhibit a peak abundance $f^{peak}_{PBH}$ exceeding $0.97$ and approaching $1$. These results indicate that a sufficient number of PBHs can form and are consistent with current observational limit. Since the corresponding density ratio $\Omega_{PBH} / \Omega_{DE}$ is approximately $0.33$, PBHs can account for one-third of the dark matter.

\subsection{Gravitational waves}

\begin{figure*}[htp]
\begin{center}
\includegraphics[width=0.50\textwidth]{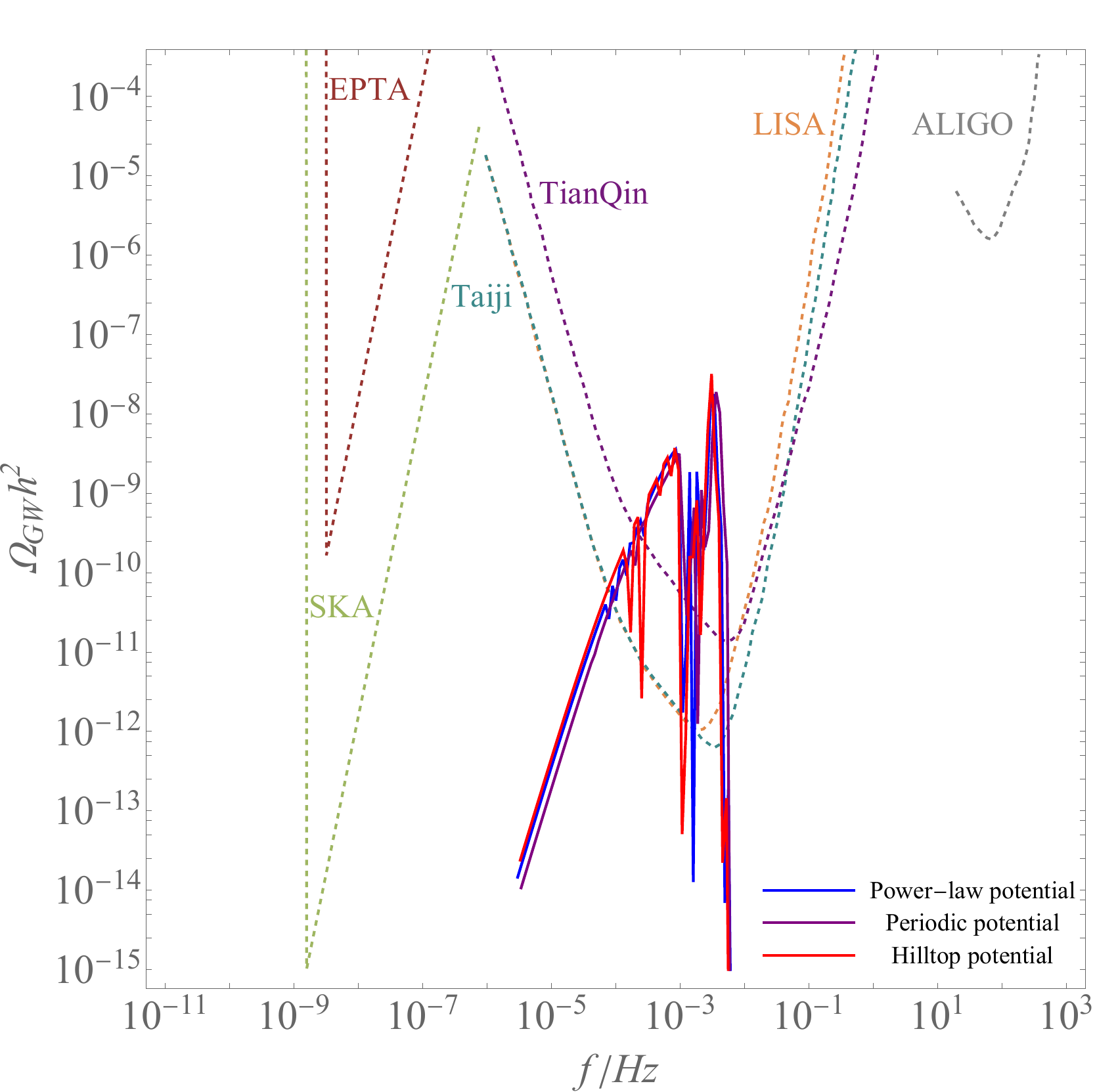}
\caption{\label{Fig70} Current energy spectrum of SIGWs.}
\end{center}
\end{figure*}

In the radiation-dominated era, large primordial scalar perturbations responsible for the formation of PBHs, generating the scalar induced gravitational waves (SIGWs) that constitute a significant observable gravitational wave (GW) signal. These SIGWs are characterized by second-order tensor perturbations $h_{ij}$, which satisfy the equation~\cite{Ananda2007, Baumann2007}
\beq\label{tp}
h''_{ij} + 2\mathcal{H}h'_{ij} - \nabla^{2}h_{ij} = -4 \mathcal{T}^{lm}_{ij}S_{lm},
\eeq
with
\beq
S_{lm} = 4 \Psi \partial_{l}\partial_{m}\Psi + 2\partial_{l}\Psi\partial_{m}\Psi - \frac{1}{\mathcal{H}^{2}}\partial_{l}(\mathcal{H}\Psi+\Psi')\partial_{m}(\mathcal{H}\Psi+\Psi'),
\eeq
where $\mathcal{T}^{lm}_{ij}$ is the transverse-traceless projection operator, and $S_{lm}$ is the GW source term. During the radiation-dominated epoch, the evolution of the metric scalar perturbation $\Psi$ is described by~\cite{Baumann2007}
\beq
\Psi_{k}(\eta) = \psi_{k} \frac{9}{(k\eta)^{2}}\Big[ \frac{\sin\big(\frac{k\eta}{\sqrt{3}}\big)}{\frac{k\eta}{\sqrt{3}}}-\cos\Big(\frac{k\eta}{\sqrt{3}}\Big) \Big],
\eeq
where $\psi_{k}$ denotes the primordial perturbation, which relates to the power spectrum of the primordial curvature perturbations through
\beq
\langle \psi_{k} \psi_{\tilde{k}} \rangle = \frac{2\pi^{2}}{k^{3}}\Big(\frac{4}{9}\mathcal{P}_{\mathcal{R}}(k)\Big)\delta(k+\tilde{k}).
\eeq
The GW energy density for each logarithmic interval $k$ can be obtained by solving Eq.~(\ref{tp}) and is formulated as~\cite{Kohri2018}
\bea\label{OGW}
\Omega_{GW}(\eta_{c},k) && = \frac{1}{12} \int^{\infty}_{0}dv \int^{\mid 1+v \mid}_{\mid 1-v \mid}du \Big[ \frac{4v^{2}-(1+v^{2}-u^{2})^{2}}{4uv} \Big]^{2} \mathcal{P}_{\mathcal{R}}(ku) \mathcal{P}_{\mathcal{R}}(kv)\nonumber\\
&& \times \Big( \frac{3}{4u^{3}v^{3}} \Big)^{2} (u^{2}+v^{2}-3)^{2}\nonumber\\
&& \times \Big[\Big( -4uv+(u^{2}+v^{2}-3)\ln \Big| \frac{3-(u+v)^{2}}{3-(u-v)^{2}} \Big| \Big)^{2} \nonumber\\
&& + \pi^{2}(u^{2}+v^{2}-3)^{2} \Theta (u+v-\sqrt{3}) \Big],
\eea
where $\Theta$ denotes the Heaviside step function, and $\eta_{c}$ represents the time when $\Omega_{GW}$ stops growing. Then, the current energy density spectrum of SIGWs can be expressed as~\cite{Kohri2018}
\beq\label{OGW0}
\Omega_{GW,0}h^{2} = 0.83\big(\frac{g_{*}}{10.75}\big)^{-\frac{1}{3}} \Omega_{r,0}h^{2}\Omega_{GW}(\eta_{c},k).
\eeq
Here, $\Omega_{r,0}h^{2}$ is the radiation density parameter at the present epoch and is set to $4.2 \times 10^{-5}$. The observational frequency $f$ of SIGWs today relates to $k$ through
\beq\label{OGWf}
f = 1.546 \times 10^{-15} \frac{k}{1\mathrm{Mpc^{-1}}}\mathrm{Hz}.
\eeq
Combining Eqs.~(\ref{OGW}), ~(\ref{OGW0}), and ~(\ref{OGWf}) with the numerical results of the primordial curvature perturbation power spectra $\mathcal{P_{R}}$ in the left panel of Fig.~(\ref{Fig6}), we calculate the current energy spectra of SIGWs and show them in Fig.~(\ref{Fig70}). In this figure, we overlay our numerical results with the sensitivity bands of several gravitational wave detectors, including Square Kilometer Array (SKA~\cite{Carilli2004}), EPTA~\cite{Lentati2015}, Taiji~\cite{HuW2017}, TianQin~\cite{Luo2016}, Laser Interferometer Space Antenna (LISA~\cite{Amaro-Seoane2017}), and Advanced Laser Interferometer Gravitational-Wave Observatory (ALIGO~\cite{Aasi2015}). This figure demonstrates that the current energy spectra of SIGWs from the power-law, periodic, and hilltop potential models exhibit a multi-peak structure, and that the corresponding peak frequencies of SIGWs are on the order of $10^{-3} \mathrm{Hz}$, as shown in Tab.~(\ref{Tab1}). Significantly, these spectra exceed the sensitivity thresholds of LISA, Taiji, and TianQin, indicating that the predicted GW signals could be detected by next-generation missions including LISA, Taiji, and TianQin.

\subsection{Evolution}

In the second previous subsection, we analyzed the formation of PBHs in constant-roll inflation within the Barrow entropy model and found that the mass of the formed PBHs is nearly $10^{-12}M_{\odot}$. In this subsection, we examine the evolution of PBHs assuming their formation occurred during the radiation-dominated era within Barrow entropy model. The black holes (BHs) can evolve with an increasing mass by absorbing other matter, stars and BHs. During this process, BHs can emit particles. The quantum properties of BHs show that the possibility of emitting particles with a thermal spectrum is related to BHs' surface gravity~\cite{Hawking1974}. During the process of emitting particles, BHs may lose mass. In the following, we will analyze the thermal properties of evaporating PBHs. The corresponding PBHs temperatures are given as~\cite{Hawking1974, Coogan2021}

\beq\label{TBH}
T_{PBH}=\frac{\hbar c^{3}}{8\pi G M_{PBH} k} \sim 1.06 \Big(\frac{10^{10}}{M_{PBH}}\Big)GeV,
\eeq
where $M_{PBH}$ is the total mass of PBHs.

To discuss the evaporation mass of PBHs, we consider Hawking evaporation to be the main process responsible for decreasing the mass of PBHs, which is defined as~\cite{Page1976a, Nayak2011}
\beq\label{dM}
\Big( \frac{dM}{dt} \Big)_{eva}=-\frac{\hbar c^{4}}{G^{2}}\frac{\alpha_{s}}{M^{2}},
\eeq
where $\alpha_{s}$ is the spin parameter of the emitting particle. Integrating Eq.~(\ref{dM}), the evaporation mass of PBHs is obtained
\beq\label{Meva}
M_{eva}=M_{i}\Big(1-\frac{t}{t_{eva}}\Big)^{\frac{1}{3}},
\eeq
with
\beq\label{teva}
t_{eva}=\frac{G^{2}}{\hbar c^{4}} \frac{M^{3}_{i}}{3\alpha_{s}},
\eeq
where $M_{i}$ is the initial mass of PBHs and $t_{eva}$ denotes the Hawking evaporation time scale. The initial mass of PBHs is given as the order of the particle horizon mass when it was formed~\cite{Jamil2011}
\beq\label{tf}
M_{i}\approx \frac{c^{3}t_{f}}{G} \approx \frac{10^{12}t_{f}}{10^{-23}}.
\eeq
Here, $t_f$ is the time of its formation. So, the PBHs formed in the late time of the universe must have more mass than that formed in the early time. For the case of PHBs formed at Planck time $10^{-43}s$, it had the mass $10^{-8}kg$. For an initial mass $M_{i}=2 \times 10^{30}kg$, which is the mass of sun $M_{\odot}$, it requires $t_{f}=2 \times 10^{-5}s$. For $M_{i}=1.1 \times 10^{-12} M_{\odot}$, which is the peak masse of PBHs obtained in previous subsection, it requires $t_{f}=2.2 \times 10^{-17}s$.

\begin{figure*}[htp]
\begin{center}
\includegraphics[width=0.40\textwidth]{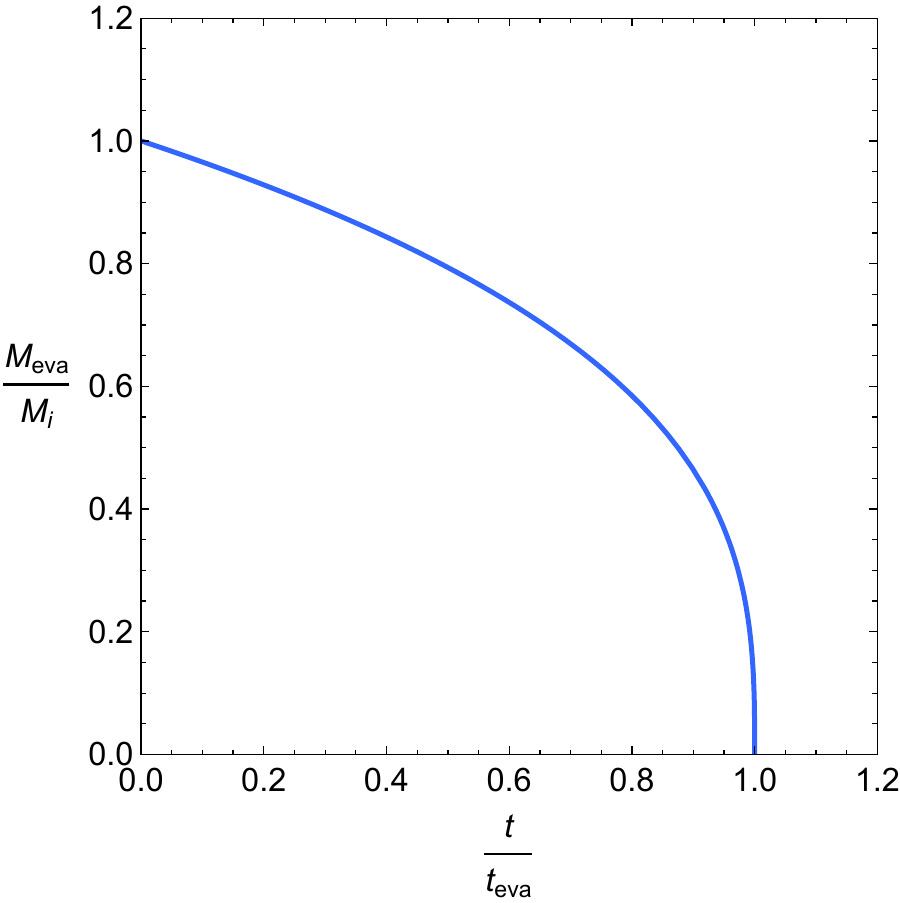}
\includegraphics[width=0.46\textwidth]{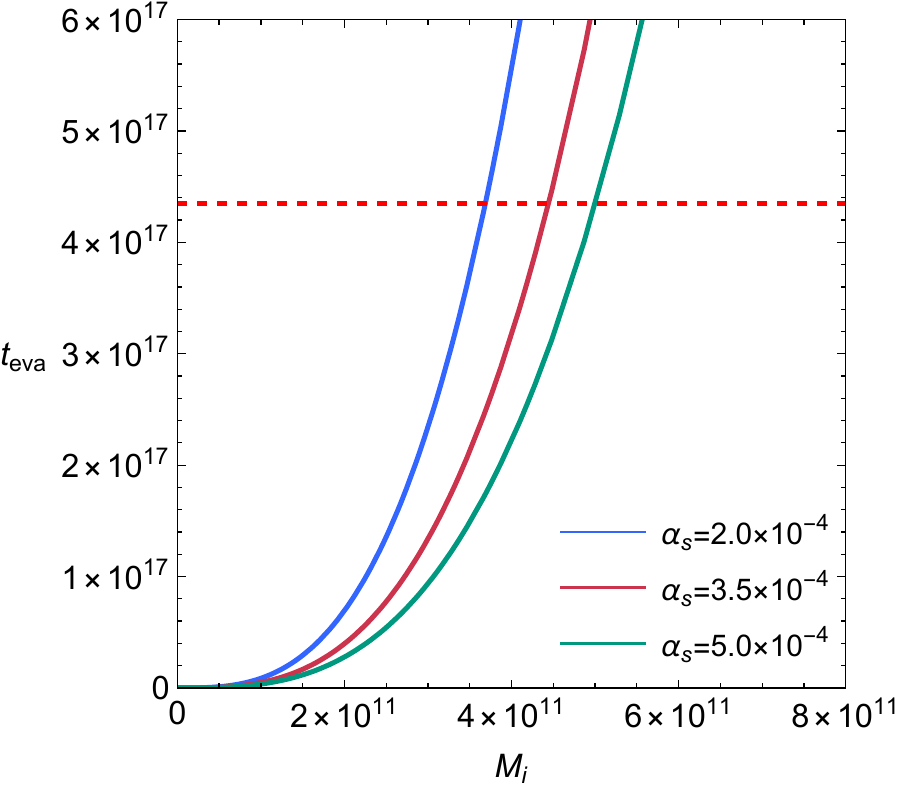}
\caption{\label{Fig7} The evolution of $M_{eva}/M_{i}$ as a function of $t/t_{eva}$ is plotted in the left panel, and the relation between the initial mass of PBHs $M_{i}$ and the evaporation time $t_{eva}$ is shown in the right panel. The red dashed line denotes the age of the universe.}
\end{center}
\end{figure*}

Eq.~(\ref{Meva}) shows the evolution of the evaporation mass of PBHs with respect to $t/t_{eva}$. The evaporation mass decreases as $t$ approaches to $t_{eva}$ and becomes $0$ for $t=t_{eva}$, which indicates PBHs evaporate completely. The evolution of $M_{eva}/M_{i}$ as the functions of $t/t_{eva}$ is plotted in the left panel of Fig.~(\ref{Fig7}), which is also given in Ref.~\cite{Bourakadi2023}. So, the Hawking evaporation time scale $t_{eva}$ determines the evaporation time of PBHs. According to Eq.~(\ref{teva}), we can find that $t_{eva}$ increases with the increase of $M_{i}$ and decreases with the increase of $\alpha_{s}$. In the right panel of Fig.~(\ref{Fig7}), by considering the evaporation time $t_{eva}$ as a function of the initial mass of PBHs $M_{i}$, we have plotted the relation between $t_{eva}$ and $M_{i}$. The red dashed line in this figure represents the current age of the universe. This figure shows that the PBHs need more time to achieve a complete evaporation with the increase of the initial mass, and it needs an evaporation time longer than the age of the universe when the initial mass is larger than $5 \times 10^{11}kg$, which is shown in Ref.~\cite{Page1976}. For $\alpha_{s} \sim 10^{-4}$~\cite{Page1976a, Jamil2011}, we can write $t_{eva}$ as
\beq\label{teva2}
t_{eva} \sim 10^{-17} M^{3}_{i}.
\eeq
For the initial mass equal to the peak mass of PBHs $M_{i}=1.1 \times 10^{-12} M_{\odot}=2.2 \times 10^{18}kg$, the evaporation time $t_{eva}$ is nearly $10^{38}s$, which is much larger than the age of the universe. So, in these models, PBHs have not evaporated completely at the present time.

\begin{figure*}[htp]
\begin{center}
\includegraphics[width=0.435\textwidth]{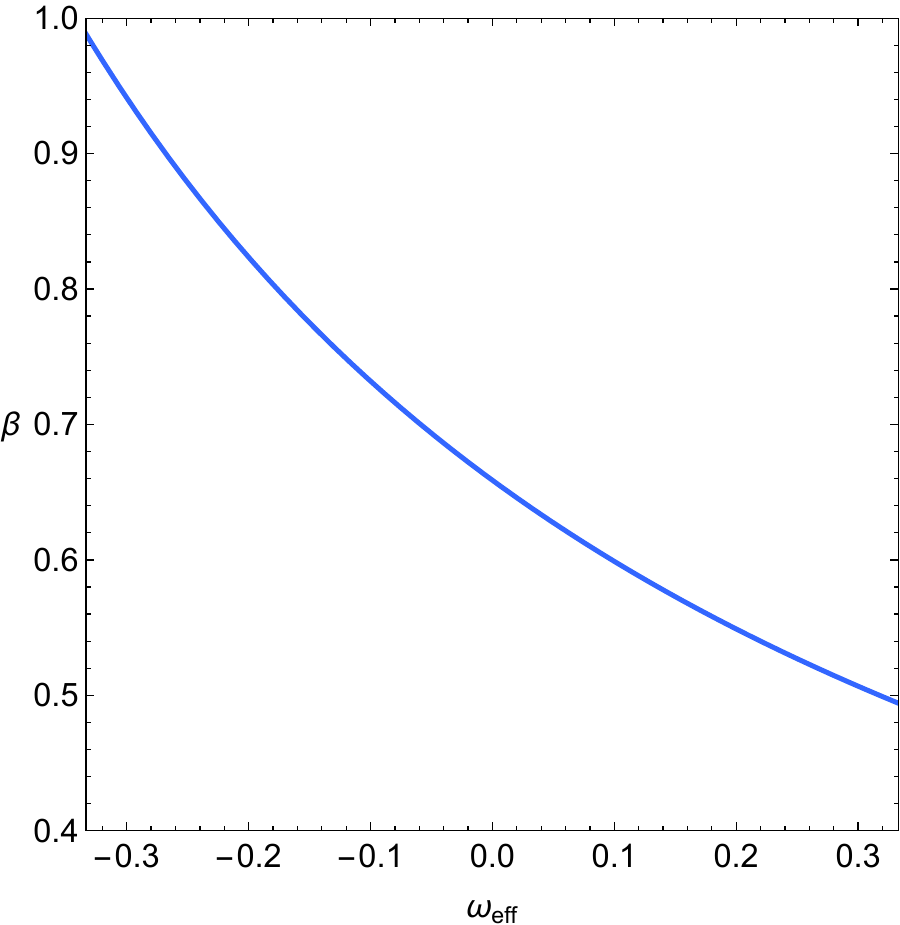}
\includegraphics[width=0.46\textwidth]{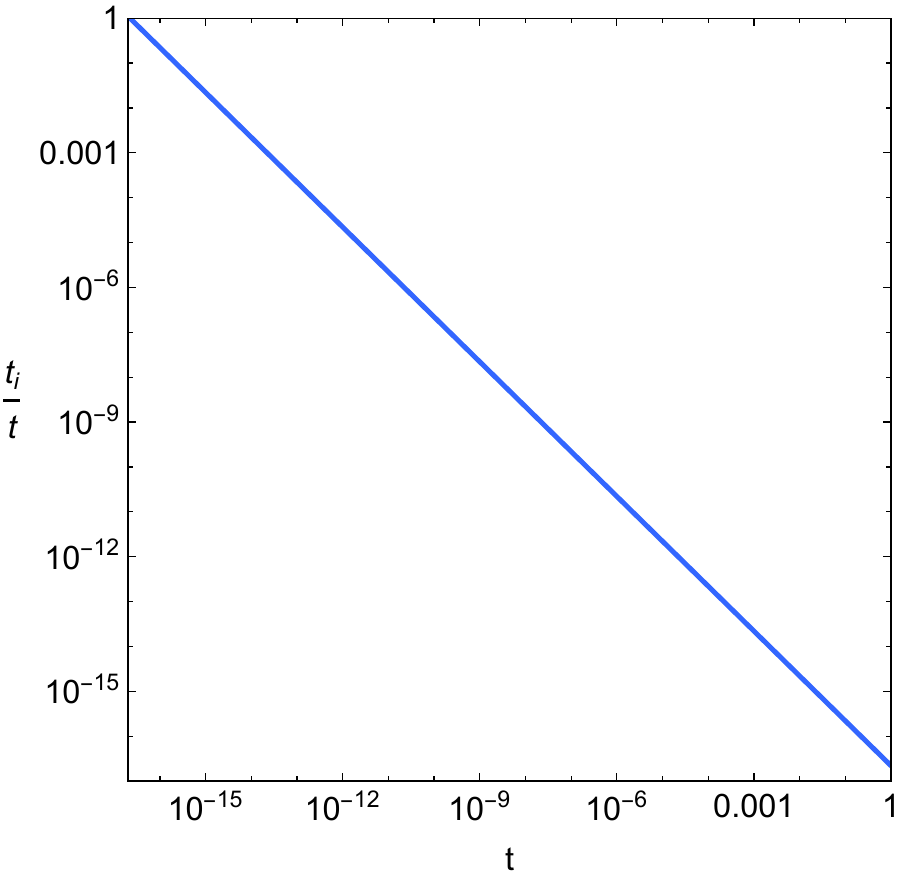}
\caption{\label{Fig8} The evolution of $\beta$ as a function of $\omega_{eff}$ is plotted in the left panel, and that of $t_{i}/t$ as a function of $t$ is plotted in the right panel. The initial mass of PBHs takes the value $M_{i}=1.1 \times 10^{-12}M_{\odot}$.}
\end{center}
\end{figure*}

\begin{figure*}[htp]
\begin{center}
\includegraphics[width=0.445\textwidth]{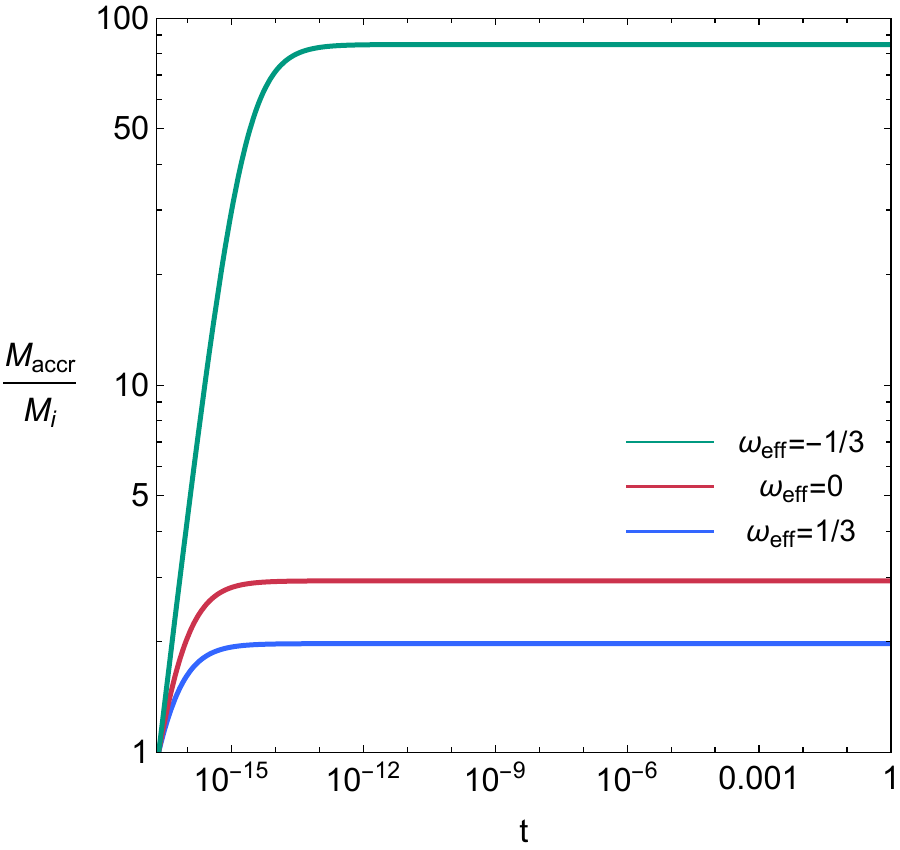}
\includegraphics[width=0.455\textwidth]{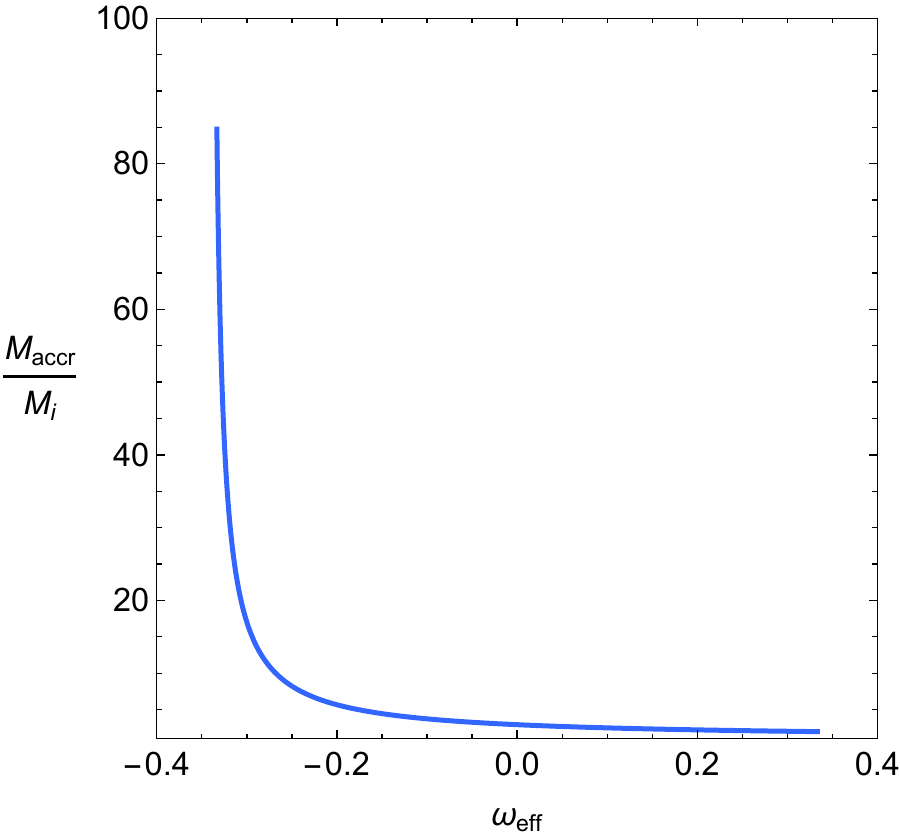}
\caption{\label{Fig9} Evolutionary curves of $M_{accr}/M_{i}$. The left panel sets $t$ as the variable, while the right panel sets $\omega_{eff}$. The initial mass of PBHs takes the value $M_{i}=1.1 \times 10^{-12}M_{\odot}$.}
\end{center}
\end{figure*}

During the process of evaporation, the accretion of fluid surrounding PBHs will prolong the evaporation of PBHs. So, we require to consider the process of the mass accretion rate for PBHs with fluid, which is given as~\cite{Babichev2004, Jamil2011}
\beq\label{dMdt}
\Big(\frac{dM}{dt}\Big)_{accr}=\frac{16\pi G^{2}}{c^{3}} M^{2}(\rho_{eff}+p_{eff}).
\eeq
Here, $\rho_{eff}$ and $p_{eff}$ represent the effective energy density and pressure, respectively. Using Eqs.~(\ref{H1}) and ~(\ref{H2}), we can obtain
\beq
\rho_{eff} = \rho_{\phi}H^{\delta} \simeq \frac{3}{8\pi G}H^{2}, \qquad \omega_{eff}=\frac{p_{eff}}{\rho_{eff}}=\frac{2\omega_{\phi}+\delta}{2-\delta},
\eeq
in which $\omega_{\phi}=p_{\phi}/\rho_{\phi}$. When $\delta$ takes a small value, one can obtain $\omega_{eff} \approx \omega_{\phi}$. Considering $a \sim t^{\frac{2}{3(1+\omega_{eff})}}$ and $H=\frac{2}{3(1+\omega_{eff}t)}$, we can write $\rho_{eff}+p_{eff}$ as
\beq
\rho_{eff}+p_{eff}=\frac{1}{6\pi G (1+\omega_{eff}) t^{2}}.
\eeq
Here, we consider that $\rho_{eff}+p_{eff}$ evolves with $t$ instead of a constant in Ref.~\cite{Bourakadi2023}. Then, Eq.~(\ref{dMdt}) can be integrated as
\beq\label{Maccr1}
M_{accr}=\frac{M_{i}}{1-\beta\Big(1-\frac{t_{i}}{t}\Big)}
\eeq
with
\beq
\beta=\frac{8G}{3c^{2}(1+\omega_{eff})}\frac{M_{i}}{t_{i}},
\eeq
in which $t_{i}$ is the time that PBHs begins to accrete, $\beta$ denotes the product of the accretion efficiency and the fraction of the horizon mass~\cite{Nayak2010}. Assuming PBHs begins to accrete at the time when it formed, we obtain $t_{i}=t_{f}$. Then, using Eq.~(\ref{tf}), $\beta$ and $t_{i}/t$ can be written as
\beq
\beta \approx \frac{0.6588}{1+\omega_{eff}},
\eeq
which indicates $\beta$ is only determined by $\omega_{eff}$ and decreases with the increase of $\omega_{eff}$, and
\beq
\frac{t_{i}}{t}=\frac{10^{-35} M_{i}}{t},
\eeq
which equals to $1$ at the time PBHs begins to accrete and decays very fast. So, according to Eq.~(\ref{Maccr1}), we can obtain $M_{accr} \simeq M_{i}$ at the initial time $t_{i}$. The evolution curve of $\beta$ and $t_{i}/t$ are plotted in the left and right panel of Fig.~(\ref{Fig8}) respectively. In the left panel of Fig.~(\ref{Fig9}), we have plotted the evolutionary curves for $M_{accr}/M_{i}$ as the function of $t$. This figure shows that $M_{accr}/M_{i}$ increases with the decrease of $\omega_{eff}$, and it increases rapidly in a short time and then keeps as a constant. The right panel of Fig.~(\ref{Fig9}) shows the relation between $M_{accr}/M_{i}$ and $\omega_{eff}$. From these figures, we can see that $M_{accr}/M_{i}$ increases to approach $10^{2}$ when $\omega_{eff}$ decreases from $1/3$ to $-1/3$.

\begin{figure*}[htp]
\begin{center}
\includegraphics[width=0.46\textwidth]{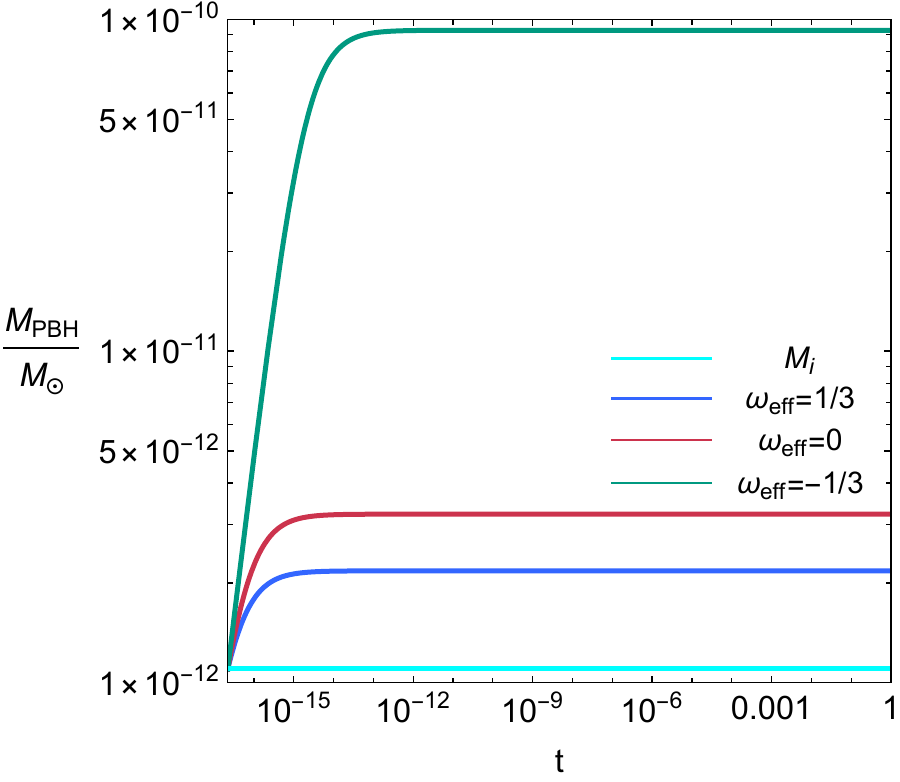}
\includegraphics[width=0.45\textwidth]{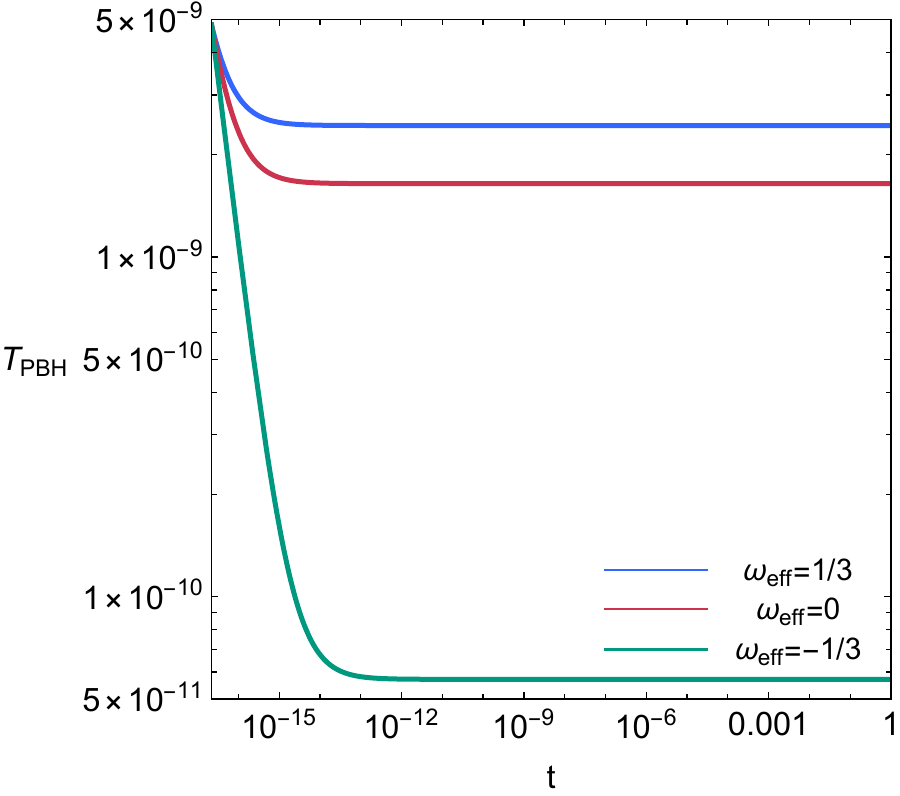}
\caption{\label{Fig10} Evolutionary curves of PBH mass $M_{PBH}$ and PBH temperature $T_{PBH}$. The initial mass of PBHs takes the value $M_{i}=1.1 \times 10^{-12}M_{\odot}$.}
\end{center}
\end{figure*}

Considering PBHs begin to accrete and evaporate after their formation, we plot the evolutionary curves of PBH mass $M_{PBH}$ and temperature $T_{PBH}$ in Fig.~(\ref{Fig10}). The left panel of Fig.~(\ref{Fig10}) shows the evolutionary curves of PBHs mass $M_{PBH}$ as the function of $t$. From this figure, we can see that $M_{PBH}$ increases rapidly in a short time and then keeps as a constant, and $M_{PBH}$ increases with the decrease of $\omega_{eff}$. In the right panel of Fig.~(\ref{Fig10}), we have plotted the evolutionary curves of PBHs temperature $T_{PBH}$ as the function of $t$. Since $T_{PBH}$ is inversely proportional to $M_{PBH}$, which is given in Eq.~(\ref{TBH}), $T_{PBH}$ decreases rapidly in a very short time and then becomes a constant, and $T_{PBH}$ decreases with the decrease of $\omega_{eff}$. When $\omega_{eff}$ evolves from $1/3$ to $-1/3$, the PBHs temperature $T_{PBH}$ decreases from $10^{-9}GeV$ to $10^{-11}GeV$, i.e. $T_{PBH}$ decreases from $10^{4}K$ to $10^{2}K$. These results indicate that PBHs still exist and can be detected today, but they persist at relatively low temperatures.

After inflation ended and PHBs formed, the universe evolves from the radiation dominated epoch to the pressureless matter dominated epoch, and then enters into the dark energy dominated epoch, the effective equation of state parameter evolves from $1/3$ to less than $-1/3$, the accretion mass $M_{accr}$ increases to approach $10^{2}M_{i}$, the PBHs temperature $T_{PBH}$ decreases with the increase of $M_{PBH}$.

\section{Conclusion}

Based on the holographic principle and Barrow entropy, a new model with the apparent horizon as an IR cutoff has been proposed, and the corresponding Friedmann equation has been modified. In this paper, starting from the modified Einstein field equations, we derive the modified scalar spectral index $n_{s}$ and the modified tensor-to-scalar ratio $r$ in Barrow entropy model. Using the modified Friedmann equation and the constant-roll condition, we calculate the constant-roll parameters $\epsilon_{1}$ and $\epsilon_{2}$, the scalar spectral index $n_{s}$, and the tensor-to-scalar ratio $r$ for the power-law, periodic, and hilltop potential models. By applying the Planck 2018 data, we plot the $r_{0.002}-n_{s}$ plane to obtain suitable values for the model parameters $\delta$. 
We find that the case of $\delta=10^{-4}$ is indistinguishable from $\delta=0$, and an increasing value of $\delta$ has a small effect on $n_{s}$ but causes a significant decrease in $r$. The value of $\gamma$ is very small, and a larger value of $\delta$ expands the range of the parameter plane. The model parameter $\delta$ and the constant-roll parameter $\gamma$ have a significant influence on $(r_{0.002},n_{s})$ behavior, exhibiting pronounced deviations from slow-roll inflation predictions. When $\delta=10^{-4}$, $n_{s}=0.9668 \pm 0.0037$ and $r_{0.002}<0.058$ are taken into consideration, the power-law potential model indicates an upper bound of $n \leq 1.46$; the periodic potential model is constrained by observed data to $f \leq 9.17$; the hilltop potential model is observationally supported for the analyzed cases of $p=1,2,3,4$. For $\gamma=0.012$, Planck 2018 data supports the $\delta$ range $0 \leq \delta \leq 0.7$ in the power-law potential model with $n=1$, $0 \leq \delta \leq 0.2$ in the periodic potential model with $f=7$, and $0 \leq \delta \leq 0.7$ in the hilltop potential model with $p=1$ and $\mu=20$, whereas the hilltop potential model with $p=4$ and $\mu=150$ reduces the supported range to $0 \leq \delta \leq 0.5$. Through application of the constant-roll condition, we derive analytical attractor solutions for these potentials and numerically verify the convergence of phase-space trajectories with disparate initial conditions to these solutions.

Then, we analyze the formation and evolution of PBHs within the Barrow entropy model, which formed during the radiation-dominated era following the end of inflation. By considering the results of constant-roll inflation and the parametric resonance mechanisms for the power-law, periodic, and hilltop potential models, we calculate the primordial curvature perturbation power spectra of these models and obtain a peak amplitude of the order of $10^{-2}$ at small scales. Based on these spectra, we investigate the abundance of PBHs and obtain the PBH mass approximately as $10^{-12} M_{\odot}$ with a peak abundance $f^{peak}_{PBH} > 0.97$ and the corresponding density ratio $\Omega_{PBH} / \Omega_{DM}$ is approximately $0.33$. These results indicate that such models both generate sufficient PBHs and account for approximately one-third of the dark matter content. Using the numerical results of primordial curvature perturbation power spectra $\mathcal{P_{R}}$, we calculate the current energy spectra of SIGWs and obtain their peak frequencies to be the order of $10^{-3} \mathrm{Hz}$. These spectra exceed the sensitivity thresholds of LISA, Taiji, and TianQin, indicating that the predicted GW signals could be detected by next-generation missions such as LISA, Taiji, and TianQin. Subsequently, by analyzing the evolution of the evaporation mass $M_{eva}$ and the accretion mass $M_{accr}$ of PBHs with the initial mass $M_{i}=10^{-12} M_{\odot}$ when the effective equation of state parameter $\omega_{eff}$ evolves from $1/3$ to $-1/3$, we find the accretion mass $M_{accr}$ increases to approximately $10^{2}M_{i}$, while the temperature of PBHs $T_{PBH}$ decreases from $10^{4}K$ to $10^{2}K$. These results indicate that PBHs still exist and can be detected today, but they persist at relatively low temperatures.

\begin{acknowledgments}

This work was supported by the National Natural Science Foundation of China under Grants Nos.12405081, 12265019, 11865018, 12305056, 11505004.

\end{acknowledgments}


\begin{thebibliography}{00}

\bibitem{Guth1981} A. Guth, Phys. Rev. D {\bf 23}, 347 (1981).

\bibitem{Linde1982} A. Linde, Phys. Lett. B {\bf 108}, 389 (1982).

\bibitem{Mukhanov1981} V. Mukhanov, G. Chibisov, JETP Lett. {\bf 33}, 532 (1981).

\bibitem{Lewis2000} A. Lewis, A. Challinor, A. Lasenby, Astrophys. J. {\bf 538}, 473 (2000).

\bibitem{Bernardeau2002} F. Bernardeau, S. Colombi, E. Gaztanaga, R. Scoccimarro, Phys. Rep. {\bf 367}, 1 (2002).

\bibitem{Weinberg2008} S. Weinberg, Cosmology, Oxford Univ. Press (2008).

\bibitem{Planck2020} Planck Collaboration, A$\&$A {\bf 641}, A10 (2020).

\bibitem{Maggiore2018} M. Maggiore, Gravitational Waves, Oxford Univ. Press (2018).

\bibitem{Noh2001} H. Noh and J. Hwang, Phys. Lett. B {\bf 515}, 231 (2001).

\bibitem{Martin2013} J. Martin, H. Motohashi, T. Suyama, Phys. Rev. D {\bf 87}, 023514 (2013).

\bibitem{Dimopoulos2017} K. Dimopoulos, Ultra slow-roll inflation demystified. Phys. Lett. B {\bf 775}, 262 (2017).

\bibitem{Pattison2018} C. Pattison, V. Vennin, H. Assadullahi, and D. Wands, JCAP {\bf 08}, 048 (2018).

\bibitem{Motohashi2015} H. Motohashi, A. Starobinsky, J. Yokoyama, JCAP {\bf 09}, 018 (2015).

\bibitem{Gao2017} Q. Gao, Sci. China Phys. Mech. Astron. {\bf 60}, 090411 (2017).

\bibitem{Gao2018} Q. Gao, Sci. China Phys. Mech. Astron. {\bf 61}, 070411 (2018).

\bibitem{Yi2018} Z. Yi, Y. Gong, JCAP {\bf 03}, 052 (2018).

\bibitem{Anguelova2018} L. Anguelova, P. Suranyi, L.C.R. Wijewardhana, JCAP {\bf 02}, 004 (2018).

\bibitem{Cicciarella2018} F. Cicciarella, J. Mabillard, M. Pieroni, JCAP {\bf 01}, 024 (2018).

\bibitem{Guerrero2020} M. Guerrero, D. Rubiera-Garcia, D. Gomez, Phys. Rev. D {\bf 102}, 123528 (2020).

\bibitem{Nojiri2017a} S. Nojiri, S. Odintsov, and V. Oikonomou, Class. Quantum Grav. {\bf 34}, 245012 (2017).

\bibitem{Motohashi2017} H. Motohashi, A. Starobinsky, Eur. Phys. J. C {\bf 77}, 538 (2017).

\bibitem{Panda2023} S. Panda, A. Rana, and R. Thakur, Eur. Phys. J. C {\bf 83}, 297 (2023).

\bibitem{Bourakadi2023} K. Bourakadi, M. Koussour, G. Otalora, M. Bennai, and T. Ouali, Phys. Dark Universe {\bf 41}, 101246 (2023).

\bibitem{Motohashi2019} H. Motohashi and A. Starobinsky, JCAP {\bf 11}, 025 (2019).

\bibitem{Mohammadi2020} A. Mohammadi, T. Golanbari, S. Nasri, and K. Saaidi, Phys. Rev. D {\bf 101}, 123537 (2020).

\bibitem{Lahiri2022} S. Lahiri, Mode. Phys. Lett. A {\bf 37}, 2250003 (2022).

\bibitem{Herrera2023} R. Herrera, M. Shokri, and J. Sadeghi, Phys. Dark Universe {\bf 41}, 101232 (2023).

\bibitem{Karam2018} A. Karam, L. Marzola, T. Pappas, A. Racioppi, and K. Tamvakis, JCAP {\bf 05}, 011 (2018).

\bibitem{Shokri2021} M. Shokri, J. Sadeghi, M. Setare, and S. Capozziello, Int. J. Mode. Phys. D {\bf 30}, 2150070 (2021).

\bibitem{Liu2024} J. Liu, Y. Gong, Z. Yi, Commun. Theor. Phys. {\bf 76} 095401 (2024).

\bibitem{Shokri2022} M. Shokri, J. Sadeghi, S. Gashti, Phys. Dark Universe {\bf 35} 100923 (2022).

\bibitem{Mohammadi2022} A. Mohammadi, Phys. Dark Universe {\bf 36}, 101055 (2022).

\bibitem{Nojiri2023} S. Nojiri, S. Odintsov, and T. Paul, Phys. Lett. B {\bf 841}, 137926 (2023).

\bibitem{Keskin2023} A. Keskin and K. Kurt, Eur. Phys. J. C {\bf 83}, 72 (2023).

\bibitem{Hawking1971} S. Hawking, Mon. Not. Roy. Astron. Soc. {\bf 152}, 75 (1971).

\bibitem{Carr1974} B. Carr and S. Hawking, Mon. Not. Roy. Astron. Soc. {\bf 168}, 399 (1974).

\bibitem{Chapline1975} G. Chapline, Nature {\bf 253}, 251 (1975).

\bibitem{Bird2016} S. Bird, I. Cholis, J. Munoz, Y. Ali-Haimoud, M. Kamionkowski, E. Kovetz, A.Raccanelli, and A. Riess, Phys. Rev. Lett. {\bf 116}, 201301 (2016).

\bibitem{Sasaki2016} M. Sasaki, T. Suyama, T. Tanaka, and S. Yokoyama, Phys. Rev. Lett. {\bf 117}, 061101 (2016).

\bibitem{Abbott2016} B. Abbott $\textit{et al}$. (LIGO Scientific Collaboration and Virgo Collaboration), Phys. Rev. D {\bf 93}, 122003 (2016).
    
\bibitem{Bagui2025} E. Bagui, S. Clesse, V. Luca, \textit{et.al.}, Living Rev. Relativ. {\bf 28}, 1 (2025).

\bibitem{Barrow1996} J. Barrow and B. Carr, Phys. Rev. D {\bf 54}, 3920 (1996).

\bibitem{Bhadra2013} J. Bhadra and U. Debnath, Int. J. Theor. Phys. {\bf 53}, 645 (2013).

\bibitem{Aliferis2021} G. Aliferis and V. Zarikas, Phys. Rev. D {\bf 103}, 023509 (2021).

\bibitem{Chanda2022} A. Chanda and B. Paul, Eur. Phys. J. C {\bf 82}, 616 (2022).

\bibitem{Bourakadi2022} K. Bourakadi, B. Asfour, Z. Sakhi, M. Bennai, and T. Ouali, Eur. Phys. J. C {\bf 82}, 792 (2022).

\bibitem{Carr1993} B. Carr and J. Lidsey, Phys. Rev. D {\bf 48}, 543 (1993).

\bibitem{Ivanov1994} P. Ivanov, P. Naselsky, and I. Novikov, Phys. Rev. D {\bf 50}, 7173 (1994).

\bibitem{Yokoyama1998} J. Yokoyama, Phys. Rev. D {\bf 58}, 083510 (1998).

\bibitem{Bellido2017} J. Garcia-Bellido and E. Ruiz Morales, Phys. Dark Universe {\bf 18}, 47 (2017).

\bibitem{Pi2018} S. Pi, Y. Zhang, Q. Huang, and M. Sasaki, JCAP {\bf 05}, 042 (2018).

\bibitem{Biagetti2018} M. Biagetti, G. Franciolini, A. Kehagias, and A. Riotto, JCAP {\bf 07}, 032 (2018).

\bibitem{Fu2020} C. Fu, P. Wu, and H. Yu, Phys. Rev. D {\bf 102}, 043527 (2020).

\bibitem{Davies2022} M. Davies, P Carrilho, and D. Mulryne, JCAP {\bf 06}, 019 (2022).

\bibitem{Lin2020} J. Lin, Q. Gao, Y. Gong, Y. Lu, C. Zhang, and F. Zhang, Phys. Rev. D {\bf 101}, 103515 (2020).

\bibitem{Yi2021} Z. Yi, Q. Gao, Y. Gong, and Z. Zhu, Phys. Rev. D {\bf 103}, 063534 (2021).

\bibitem{Gao2021} Q. Gao, Y. Gong, and Z. Yi, Nucl. Phys. B {\bf 969}, 115480 (2021).

\bibitem{Gao2021a} Q. Gao, Sci. China Phys. Mech. Astron. {\bf 64}, 280411 (2021).

\bibitem{Yi2021a} Z. Yi, Y. Gong, B. Wang, and Z. Zhu, Phys. Rev. D {\bf 103}, 063535 (2021).

\bibitem{Wu2021} L. Wu, Y. Gong, and T. Li, Phys. Rev. D {\bf 104}, 123544 (2021).

\bibitem{Wang2024} Z. Wang, S. Gao, Y. Gong, and Y. Wang, Phys. Rev. D {\bf 109}, 103532 (2024)

\bibitem{Wang2024a} X. Wang, Y. Zhang, and M. Sasaki, JCAP {\bf 07}, 076 (2024).

\bibitem{Chen2024} L. Chen, H. Yu, and P. Wu, Phys. Lett. B {\bf 849}, 138457 (2024).

\bibitem{Solbi2024} M. Solbi and K. Karami, Eur. Phys. J. C {\bf 84}, 918 (2024).

\bibitem{Frolovsky2025} D. Frolovsky and S. Ketov, Phys. Rev. D {\bf 111}, 083533 (2025).

\bibitem{Afzal2025} A. Afzal, A. Ghoshal, and S. King, Phys. Rev. D {\bf 111}, 023050 (2025).

\bibitem{Choudhury2025} S. Choudhury, A. Karde, P. Padiyar, and M. Sami, Eur. Phys. J. C {\bf 85}, 21 (2025).

\bibitem{Cook2023} J. Cook, JCAP {\bf 07}, 031 (2023).





\bibitem{Di2018} H. Di and Y. Gong, JCAP {\bf 07}, 007 (2018).

\bibitem{Fu2019} C, Fu, P, Wu, and H, Yu, Phys. Rev. D {\bf 100}, 063532 (2019).

\bibitem{Ballesteros2020} G. Ballesteros, J. Rey, M. Taoso, and A. Urbano, JCAP {\bf 08}, 043 (2020).

\bibitem{Liu2021} Y. Liu, Q. Wang, B. Su, and N. Li, Phys. Dark Universe {\bf 34}, 100905 (2021).

\bibitem{Figueroa2022} D. Figueroa, S. Raatikainen, S. Rasanen, and E. Tomberg, JCAP {\bf 05}, 027 (2022).

\bibitem{Zhai2022} R. Zhai, H. Yu, and P. Wu, Phys. Rev. D {\bf 106}, 023517 (2022).

\bibitem{Chen2022} L. Chen, H. Yu, and P. Wu, Phys. Rev. D {\bf 106}, 063537 (2022).

\bibitem{Zhai2023} R. Zhai, H. Yu, and P. Wu, Phys. Rev. D {\bf 108}, 043529 (2023).

\bibitem{Choudhury2024} S. Choudhury, A. Karde, S. Panda, and S. SenGupta, Eur. Phys. J. C 84, 1149 (2024).

\bibitem{Su2025} B. Su, N. Li, and L. Feng, Eur. Phys. J. C {\bf 85}, 197 (2025).




\bibitem{Motohashi2020} H. Motohashi, S. Mukohyama, and M. Oliosi, JCAP {\bf 03}, 002 (2020).

\bibitem{Tomberg2023} E. Tomberg, Phys. Rev. D {\bf 108}, 043502 (2023).

\bibitem{Clesse2015} S. Clesse and J. Garcia-Bellido, Phys. Rev. D {\bf 92}, 023524 (2015).

\bibitem{Kawasaki2016} M. Kawasaki, A. Kusenko, Y. Tada, and T. T. Yanagida, Phys. Rev. D {\bf 94}, 083523 (2016).

\bibitem{Carr2016} B. Carr, F. Kuhnel, and M. Sandstad, Phys. Rev. D {\bf 94}, 083504 (2016).

\bibitem{Inomata2017} K. Inomata, M. Kawasaki, K. Mukaida, Y. Tada, and T. Yanagida, Phys. Rev. D {\bf 96}, 043504 (2017).

\bibitem{Inomata2018} K. Inomata, M. Kawasaki, K. Mukaida, and T. Yanagida, Phys. Rev. D {\bf 97}, 043514 (2018).

\bibitem{Carr2018} B. Carr and J. Silk, Mon. Not. Roy. Astron. Soc. {\bf 478}, 3756 (2018).

\bibitem{Germani2019} C. Germani and I. Musco, Phys. Rev. Lett. {\bf 122}, 141302 (2019).

\bibitem{Kusenko2020} A. Kusenko, M. Sasaki, S. Sugiyama, M. Takada, V. Takhistov, and E. Vitagliano, Phys. Rev. Lett. {\bf 125}, 181304 (2020).

\bibitem{Calabrese2022} R. Calabrese, D. Fiorillo, G. Miele, S. Morisi, and A. Palazzo, Phys. Lett. B {\bf 829}, 137050 (2022).

\bibitem{Pacheco2023} J. de Freitas Pacheco, E. Kiritsis, M. Lucca, and J. Silk, Phys. Rev. D {\bf 107}, 123525 (2023).

\bibitem{Flores2023} M. Flores and A. Kusenko, JCAP {\bf 05}, 013 (2023).

\bibitem{Dike2023} V. Dike, D. Gilman, T. Treu, Mon. Not. Roy. Astron. Soc. {\bf 522}, 5434 (2023).

\bibitem{Pacheco2020} J. de Freitas Pacheco and J. Silk, Phys. Rev. D {\bf 101}, 083022 (2020).

\bibitem{Calza2025} M. Calza, D. Pedrotti, and S. Vagnozzi, Phys. Rev. D {\bf 111}, 024010 (2025).



\bibitem{Fuller2017} G. Fuller, A. Kusenko, and V. Takhistov, Phys. Rev. Lett. {\bf 119}, 061101 (2017).

\bibitem{Keith2020} C. Keith, D. Hooper, N. Blinov, and S. McDermott, Phys. Rev. D {\bf 102}, 103512 (2020).

\bibitem{Kawasaki2012} M. Kawasaki, A. Kusenko, and T. Yanagida, Phys. Lett. B {\bf 711}, 1 (2012).

\bibitem{Wang2023} X. Wang, Y. Zhang, R. Kimura, M. Yamaguchi,  Sci. China Phys. Mech. Astron. {\bf 66}, 260462 (2023).

\bibitem{Inomata2017a} K. Inomata, M. Kawasaki, K. Mukaida, Y. Tada, and T. Yanagida, Phys. Rev. D {\bf 95}, 123510 (2017).

\bibitem{Page1976} D. Page, Phys. Rev. D {\bf 13}, 198 (1976).

\bibitem{Kuhnel2017} F. Kuhnel and K. Freese, Phys. Rev. D {\bf 95}, 083508 (2017).

\bibitem{Carr2017} B. Carr, M. Raidal, T. Tenkanen, V. Vaskonen, and H. Veermae, Phys. Rev. D {\bf 96}, 023514 (2017).

\bibitem{DAgostino2023} R. DAgostino, R. Giambo, and O. Luongo, Phys. Rev. D {\bf 107}, 043032 (2023).

\bibitem{Green2024} A. Green, Nuclear Physics B {\bf 1003}, 116494 (2024).

\bibitem{Hsu2004} S. Hsu, Phys. Lett. B {\bf 594} 13 (2004).

\bibitem{Horvat2004} R. Horvat, Phys. Rev. D {\bf 70} 087301 (2004).

\bibitem{Li2004} M. Li, Phys. Lett. B {\bf 603} 1 (2004).

\bibitem{Witten1998} E. Witten, Adv. Theor. Math. Phys. {\bf 2} 253 (1998).

\bibitem{Bousso2002} R. Bousso, Rev. Modern Phys. {\bf 74} 825 (2002).

\bibitem{Barrow2020a} J. D. Barrow, Phys. Lett. B {\bf 808} 135643 (2020).

\bibitem{Saridakis2020} E. N. Saridakis, Phys. Rev. D {\bf 102} 123525 (2020).

\bibitem{Anagnostopoulos2020a} F. K. Anagnostopoulos, S. Basilakos, and E. N. Saridakis, Eur. Phys. J. C {\bf 80} 826 (2020).

\bibitem{Srivastava2021a} S. Srivastava and U. K. Sharma, Int. J. Geom. Methods Mode. Phys. {\bf 18} 2150014 (2021).

\bibitem{Sheykhi2021} A. Sheykhi, Phys. Rev. D {\bf 103}, 123503 (2021).

\bibitem{Oliveros2022} A. Oliveros, M. Sabogal, and M. Acero, Eur. Phys. J. Plus {\bf 137}, 783 (2022).

\bibitem{Huang2021} Q. Huang, H. Huang, B. Xu, F. Tu, and J. Chen, Eur. Phys. J. C {\bf 81}, 686 (2021).

\bibitem{Adhikary2021} P. Adhikary, S. Das, S. Basilakos, and E. Saridakis, Phys. Rev. D {\bf 104}, 123519 (2021).

\bibitem{Mamon2021} A. Mamon, A. Paliathanasis, and S. Saha, Eur. Phys. J. Plus {\bf 136}, 134 (2021).

\bibitem{Rani2021} S. Rani and N. Azhar, Universe {\bf 7}, 268 (2021).

\bibitem{Luciano2022} G. Luciano, Phys. Rev. D {\bf 106}, 083530 (2022).

\bibitem{Nojiri2022} S. Nojiri, S. Odintsov, and T. Paul, Phys. Lett. B {\bf 825}, 136844 (2022).

\bibitem{Paul2022} B. Paul, B. Roy, and A. Saha, Eur. Phys. J. C {\bf 82}, 76 (2022).

\bibitem{Boulkaboul2023} N. Boulkaboul, Phys. Dark Universe {\bf 40}, 101205 (2023).

\bibitem{Pankaj2023} R. Pankaj, U. Sharma, and N. Ali, Astrophys Space Sci. {\bf 368}, 15 (2023).

\bibitem{Ghaffari2023} S. Ghaffari, G. Luciano, and S. Capozziello, Eur. Phys. J. Plus {\bf 138}, 82 (2023).

\bibitem{Luciano2023} G. Luciano, Eur. Phys. J. C {\bf 83}, 329 (2023).

\bibitem{Asghari2022} M. Asghari and A. Sheykhi, Eur. Phys. J. C {\bf 82}, 388 (2022).

\bibitem{Jusufi2022} K. Jusufi, M. Azreg-Ainou, M. Jamil, and E. Saridakis, Universe {\bf 8}, 102 (2022).

\bibitem{Barrow2021} J. Barrow, S. Basilakos, and E. Saridakis, Phys. Lett. B {\bf 815}, 136134 (2021).

\bibitem{Anagnostopoulos2020} F. Anagnostopoulos, S. Basilakos, and E. Saridakis, Eur. Phys. J. C {\bf 80}, 826 (2020).








\bibitem{Saridakis2020a} E. Saridakis, JCAP {\bf 07}, 031 (2020). 

\bibitem{Leon2021} G. Leon, J. Magana, A. Hernandez-Almada, M. Garcia-Aspeitia, T, Verdugo, and V. Motta, JCAP {\bf 12}, 032 (2021).

\bibitem{Sheykhi2022} A. Sheykhi and B. Farsi, Eur. Phys. J. C {\bf 82}, 1111 (2022).

\bibitem{Sheykhi2023} A. Sheykhi,Phys. Rev. D {\bf 107}, 023505 (2023).

\bibitem{Sheykhi2023a} A. Sheykhi and S. Ghaffari, Phys. Dark Universe {\bf 41}, 101241 (2023).

\bibitem{Salehi2023} A. Salehi, Eur. Phys. J. C {\bf 83}, 1027 (2023).

\bibitem{Motaghi2024} M. Motaghi, A. Sheykhi, and E. Ebrahimi, Phys. Dark Universe {\bf 46}, 101710 (2024).

\bibitem{Sheykhi2025} A. Sheykhi and A. Shahbazi, Phys. Rev. D {\bf 111}, 043518 (2025).

\bibitem{Keskin2025} A. Keskin, L. Canpolat, and K. Kurt, Eur. Phys. J. C {\bf 85}, 634 (2025).












\bibitem{Barrow2020} J. Barrow, Phys. Lett. B {\bf 808}, 135643 (2020).

\bibitem{Garriga1999} J. Garriga and V. Mukhanov, Phys. Lett. B {\bf 458}, 219 (1999).

\bibitem{Mukhanov1992} V. Mukhanov, H. Feldman, and R. Brandenberger, Phys. Rep. {\bf 215}, 203 (1992).

\bibitem{Handley2019} W. Handley, Phys. Rev. D 100, 123517 (2019).

\bibitem{Thavanesan2021} A. Thavanesan, D. Werth, and W. Handley, Phys. Rev. D 103, 023519 (2021). 

\bibitem{Shumaylov2022} Z. Shumaylov and W. Handley, Phys. Rev. D 105, 123532 (2022). 

\bibitem{HuangQ2022} Q. Huang, K. Zhang, Z. Fang, and F. Tu, Phys. Dark Universe 38, 101124 (2022). 

\bibitem{HuangQ2023} Q. Huang, K. Zhang, H. Huang, B. Xu, and F. Tu, Universe 9, 221 (2023).

\bibitem{HuangQ2023a} Q. Huang, H. Huang, and B. Xu, Phys. Dark Universe 41, 101262 (2023).

\bibitem{Dodelson2021} S. Dodelson and F. Schmidt, Modern Cosmology (2nd ed.). New York: Academic press, (2021).  

\bibitem{Nojiri2017} S. Nojiri, S. Odintsov, and V. Oikonomou, Phys. Rep. {\bf 692}, 1 (2017).

\bibitem{Hwang2005} J. Hwang and H. Noh, Phys. Rev. D {\bf 71}, 063536 (2005).

\bibitem{Linde1983} A. Linde, Phys. Lett. B {\bf 129}, 177 (1983).

\bibitem{Bassett2006} B. Bassett, S. Tsujikawa, and D. Wands, Rev. Mod. Phys. {\bf 78}, 537 (2006).

\bibitem{Tahmasebzadeh2016} B. Tahmasebzadeh, K. Rezazadehb, and K. Karami, JCAP {\bf 07}, 006 (2016).

\bibitem{Goswami2018} R. Goswami and U. Yajnik, JCAP {\bf 10}, 018 (2018).






\bibitem{Planck2020a} Planck Collaboration, A$\&$A {\bf 641}, A6 (2020).

\bibitem{Freese1990} K. Freese, J. Frieman, and A. Olinto, Phys. Rev. Lett. {\bf 65}, 3233 (1990).

\bibitem{Adams1993} F. Adams, J. Bond, K. Freese, J. Frieman, and A. Olinto, Phys. Rev. D {\bf 47}, 426 (1993).

\bibitem{Cook2015} J. Cook, E. Dimastrogiovanni, D. Eassona, and L. Krauss, JCAP {\bf 04}, 047 (2015).

\bibitem{Boubekeur2005} L. Boubekeur and D. Lyth, JCAP {\bf 07}, 02 (2005).

\bibitem{Remmen2013} G. Remmen and S. Carroll, Phys. Rev. D {\bf 88}, 083518 (2013).


\bibitem{Motohashi2017p} H. Motohashi and W. Hu, Phys. Rev. D {\bf 96}, 063503 (2017).

\bibitem{Young2014} S. Young, C. Byrnes, and M. Sasaki, JCAP {\bf 07}, 045 (2014).

\bibitem{Tada2019} Y. Tada and S. Yokoyama, Phys. Rev. D {\bf 100}, 023537 (2019).

\bibitem{Musco2013} I. Musco and J. Miller, Class. Quantum Gravi. {\bf 30}, 145009 (2013).

\bibitem{Harada2014} T. Harada, C. Yoo, and K. Kohri, Phys. Rev. D {\bf 88}, 084051 (2013); {\bf 89}, 029903(E) (2014).

\bibitem{Cai2020} R. Cai, Z. Guo, J. Liu, L. Liu, and X. Yang, JCAP {\bf 06}, 013 (2020).

\bibitem{Cai2018} Y. Cai, X. Tong, D. Wang, and S. Yan, Phys. Rev. Lett. {\bf 121}, 081306 (2018). 

\bibitem{Zhou2020} Z. Zhou, J. Jiang, Y. Cai, M. Sasaki, and S. Pi, Phys. Rev. D {\bf 102}, 103527 (2020).

\bibitem{Peng2021} Z. Peng, C. Fu, J. Liu, Z. Guo, and R. Cai, JCAP {\bf 10}, 050 (2021).

\bibitem{Cai2021} Y. Cai, J. Jiang, M. Sasaki, V. Vardanyan, and Z. Zhou, Phys. Rev. Lett. {\bf 127}, 251301 (2021).

\bibitem{Cai2024} Y. Cai, G. Domenech, A. Ganz, J. Jiang, C. Lin, and B. Wang, JCAP {\bf 10}, 027 (2024).

\bibitem{Cai2019} Y. Cai, C. Chen, X. Tong, D. Wang, and S. F. Yan, Phys. Rev. D {\bf 100}, 043518 (2019).

\bibitem{Chen2019} C. Chen and Y. Cai, J. Cosmol. Astropart. Phys. {\bf 10}, 068 (2019).

\bibitem{Chen2020} C. Chen, X. H. Ma and Y. F. Cai, Phys. Rev. D {\bf 102}, 063526 (2020).

\bibitem{Addazi2022} A. Addazi, S. Capozziello, and Q. Gan, J. Cosmol. Astropart. Phys. {\bf 08}, 051 (2022).

\bibitem{Yu2024} Y. Yu and S. Wang, Phys. Rev. D {\bf 109}, 083501 (2024). 

\bibitem{Fixsen1996} D. Fixsen, E. Cheng, J. Gales, J. Mather, R. Shafer, and E. Wright, Astrophys. J. {\bf 473}, 576 (1996).

\bibitem{Inomata2016} K. Inomata, M. Kawasaki, and Y. Tada, Phys. Rev. D {\bf 94}, 043527 (2016).

\bibitem{Inomata2019} K. Inomata and T. Nakama, Phys. Rev. D {\bf 99}, 043511 (2019).

\bibitem{Carr2010} B. Carr, K. Kohri, Y. Sendouda, and J. Yokoyama, Phys. Rev. D {\bf 81}, 104019 (2010).

\bibitem{Graham2015} P. Graham, S. Rajendran, and J. Varela, Phys. Rev. D {\bf 92}, 063007 (2015).

\bibitem{Laha2019} R. Laha, Phys. Rev. Lett. {\bf 123}, 251101 (2019).

\bibitem{Griest2013} K. Griest, A. Cieplak, and M. Lehner, Phys. Rev. Lett. {\bf 111}, 181302 (2013).

\bibitem{Niikura2019} H. Niikura, M. Takada, S. Yokoyama, T. Sumi, and S. Masaki, Phys. Rev. D {\bf 99}, 083503 (2019).

\bibitem{Tisserand2007} P. Tisserand, L. Le Guillou, C. Afonso, et al. (EROS-2 Collaboration), Astron. Astrophys. {\bf 469}, 387 (2007).

\bibitem{Poulin2017} V. Poulin, P. Serpico, F. Calore, S. Clesse and K. Kohri, Phys. Rev. D {\bf 96}, 083524 (2017).


\bibitem{Ananda2007} K. Ananda, C. Clarkson, and D. Wands, Phys. Rev. D {\bf 75}, 123518 (2007).

\bibitem{Baumann2007} D. Baumann, P. Steinhardt, K. Takahashi, and K. Ichiki, Phys. Rev. D {\bf 76}, 084019 (2007).

\bibitem{Kohri2018} K. Kohri and T. Terada, Phys. Rev. D {\bf 97}, 123532 (2018).



\bibitem{Carilli2004} C. Carilli and S. Rawlings, New Astron. Rev. {\bf 48}, 979 (2004).

\bibitem{Lentati2015} L. Lentati, S. Taylor, C. Mingarelli, \textit{et al.}, Mon. Not. R. Astron. Soc. {\bf 453}, 2576 (2015).

\bibitem{HuW2017} W. Hu, Y. Wu, Natl. Sci. Rev. {\bf 4}, 685 (2017).

\bibitem{Luo2016} J. Luo, L. Chen, H. Duan, \textit{et al.} (TianQin), Class. Quantum Gravity {\bf 33}, 035010 (2016).

\bibitem{Amaro-Seoane2017} P. Amaro-Seoane, H. Audley, S. Babak, \textit{et al.} (LISA), arXiv:1702.00786.

\bibitem{Aasi2015} J. Aasi, B. Abbott, R. Abbott, \textit{et al.} (The LIGO Scientific Collaboration), Class. Quantum Gravity {\bf 32}, 074001 (2015). 


\bibitem{Hawking1974} S. Hawking, Nature {\bf 248}, 30 (1974).

\bibitem{Coogan2021} A. Coogan, L. Morrison, and S. Profumo, Phys. Rev. Lett. {\bf 126}, 171101 (2021).

\bibitem{Page1976a} D. Page, Phys. Rev. D {\bf 13}, 198 (1976).

\bibitem{Nayak2011} B. Nayak and L. Singh, Pramana {\bf 76}, 173 (2011).

\bibitem{Jamil2011} M. Jamil and A. Qadir, Gen. Relativ. Gravit. {\bf 43}, 1069 (2011).

\bibitem{Babichev2004} E. Babichev, V. Dokuchaev, and Y. Eroshenko, Phys. Rev. Lett. {\bf 93}, 021102 (2004).

\bibitem{Nayak2010} B. Nayak and L. Singh, Phys. Rev. D {\bf 82}, 127301 (2010).

\end{thebibliography}
\end{document}